\documentclass[useAMS,usenatbib]{mn2e}
\usepackage{graphicx}
\usepackage{amsmath}
\usepackage{tabularx}
\usepackage{inputenc}
\usepackage{float}
\graphicspath{{images/}}
\title[Rotation periods from time-series spectroscopy]{Rotation periods of late-type dwarf stars from time-series high-resolution  spectroscopy of chromospheric indicators}

\author[A. Su\'{a}rez Mascare\~{n}o]{A. Su\'{a}rez Mascare\~{n}o$^{1,2}$ \thanks{E-mail: asm@iac.es}, R. Rebolo$^{1,2,3}$, J.~I. Gonz\'alez Hern\'andez$^{1,2}$, M. Esposito$^{1,2}$
\\
$^{1}$ Instituto de Astrof\'{i}sica de Canarias, E-38205 La Laguna, Tenerife, Spain\\
$^{2}$ Universidad de La Laguna, Dpto. Astrof�sica, E-38206 La Laguna, Tenerife, Spain\\
$^{3}$ Consejo Superior de Investigaciones Cient{\'\i}ficas, Spain\\
}

\begin{document}

\date{Written - 2015}

\pagerange{\pageref{firstpage}--\pageref{lastpage}} \pubyear{2002}

\maketitle

\label{firstpage}

\begin{abstract}

We determine rotation periods of a sample of 48  late F-type to mid-M dwarf stars  using time-series high-resolution  spectroscopy  of  the Ca II H$\&$K and H$\alpha$ chromospheric activity indicators. We find good agreement between the  rotation periods obtained from each of these two  indicators.  An empirical relationship between the level of  chromospheric emission  measured by $\log_{10}(R'_\textrm{HK}$) and the  spectroscopic rotation periods is reported. This relation is largely  independent of  the spectral type and the metallicity of the stars and  can be used to make a reliable prediction of  rotation periods for  late K to mid-M dwarfs with  low levels of activity. For some  stars in the sample, the  measured  spectroscopic rotation periods coincide, or are very close, to the orbital periods of  postulated  planets. In such cases, further studies   are needed to clarify whether the associated  periodic radial velocity signals  reveal   the existence of planets or are due to  magnetic activity.

\end{abstract}


\section{Introduction}

 Magnetic activity  influences spectral lines \citep{Dravins1985} and produces inhomogeneities in the stellar surface, which, combined with  rotation, causes  Doppler shifts of  the stellar spectrum \citep{Saar1997}. Changes in the distribution of spots induce apparent Doppler shifts from less than one to dozens of metres per second, depending on the level of stellar activity \citep{Huelamo2008, Pont2011, Hatzes2013}. When these induced signals are periodic, they could be mistaken for planetary signals.  In fact, stellar activity is  one of the most important causes of false alarm signals  in the analysis of high precision RV time series. Disentangling these signals from a real planetary signal in RV curves  is a crucial issue  when searching for terrestrial exoplanets.  Since the exoplanet community started to search for small planets there has always been a concern that the intrinsic variability of the stellar surface \citep{Baliunas1997} could render impossible the detection of the low amplitude radial velocity (RV) signals caused by terrestrial planets \citep{QuelozHenry2001,  DumusquePepe2012, Santos2014}. The precise  determination of the rotation periods of stars is a crucial first step  in discriminating true planetary signals from those induced by the stellar activity.

A well-known connection between stellar activity and rotation has been widely used to estimate rotation periods through stellar models \citep{Skumanich1972, CatalanoMarilli1983,  Noyes1983}. Rotation periods have usually been measured from periodic variations in the stellar luminosity associated with the  distribution of superficial  magnetic spots. These surface structures  are  sufficiently  stable in time to span  longer than the rotation period of the star  \citep{Robertson2015} and to  make possible  the measurement of stellar  rotation periods  from extensive photometry time series \citep{MessinaGuinan2002, Kiraga2007}. However, these  photometric variations are not always strong enough to be detected in stars of very low activity where many RV searches have been performed.  

Stellar induced radial velocity variations come mainly from two different sources. First, stellar spots are cooler and radiate less than other regions of the stellar photosphere, thus affecting the photospheric line profiles and therefore the RV measurements. This effect changes as the star rotates. Then stellar magnetic activity creates regions in which the convective currents are altered, thereby  changing the ingoing or outgoing flux of material in parts of the star's surface \citep{Dumusque2014}.  The position and distribution of these spots and magnetically active regions is strongly dependent on the long term magnetic cycle of the star, so it usually extends over more than a few rotational periods. The second effect produces an apparent change in the radial velocity of the star that becomes more important in later stellar types \citep{Robertson2014}, but it also results in a change in the intensity of the chromospheric emission lines. Since we integrate the whole light of the visible side of the star, the observed fluxes of the lines change  with stellar rotation.  Measuring these flux variations in spectroscopic time series of chromospheric lines can therefore reveal the rotation period of a star \citep{Dravins1981, Dravins1982, Livingston1982, Brandt1990, Donahue1996, Lindegren2003, Meunier2010, Lovis2011, Howard2014}.  This  has been possible even for stars of very low activity \citep{DumusquePepe2012,Robertson2014}. In this paper we use high quality spectroscopic time series of low-activity  late-type dwarfs to investigate the time variability of chromospheric indicators and determine rotation periods.

\section[]{Stellar Sample and spectroscopic data}

Our sample consists of 48 main-sequence stars from the solar vicinity, covering from late F-type to mid M-type stars, with apparent magnitudes $m_V$ in the range 4--13. The selected stars come from planet-hunting programs using HARPS~\citep{Mayor2003} and therefore have a bias towards low chromospheric activiy ($\log_{10}(R'_\textrm{HK}) < -4.5$). We search in the HARPS ESO public data archive for stars with spectra available in more than 20 individual observing nights as for April 2014. The sampling should be appropriate for the detection of periodic signals compatible with the rotation of low activity stars. In total, more than 6400 spectra have been analysed in this work. In Table~\ref{tab:Data_Stars} we provide information on the number of spectra, time baseline, and several properties of the stars of our sample.

HARPS is a fibre-fed high resolution echelle spectrograph installed at the 3.6 m ESO telescope in La Silla Observatory (Chile). The instrument has a resolving power $R\sim 115\,000$ over a spectral range from 378 to 691 nm and has been designed to attain very high long-term radial velocity accuracy. It is contained in a vacuum vessel to avoid spectral drifts due to temperature and air pressure variations, thus  ensuring its stability.
HARPS is equipped with its own pipeline providing extracted and wavelength-calibrated spectra, as well as RV measurements and other data products such as cross-correlation functions and their bisector profiles.

For our analysis we use the extracted order-by-order wavelength-calibrated spectra produced by the HARPS pipeline. For a given star, the change in atmospheric transparency from day to day causes variations in the  flux distribution of the recorded spectra that are particularly relevant in the blue where we intend to measure Ca II lines. In order to minimize  the effects related to these atmospheric changes  we create a spectral template for each star by de-blazing and co-adding every available spectrum and use the co-added spectrum to correct the order-by-order fluxes of the individual ones. We also   correct each spectrum for the Earth's barycentric radial velocity and the radial velocity of the star using the measurements given by the standard pipeline and re-binned the spectra into a wavelength-constant step. Using this  HARPS dataset, we expect to have high quality spectroscopic indicators  to monitor tiny stellar activity variations with  high accuracy.

\section[]{ACTIVITY INDICATORS }

\subsection{The HARPS Ca II H\&K Index}

\begin{figure}
	\includegraphics[width=\linewidth]{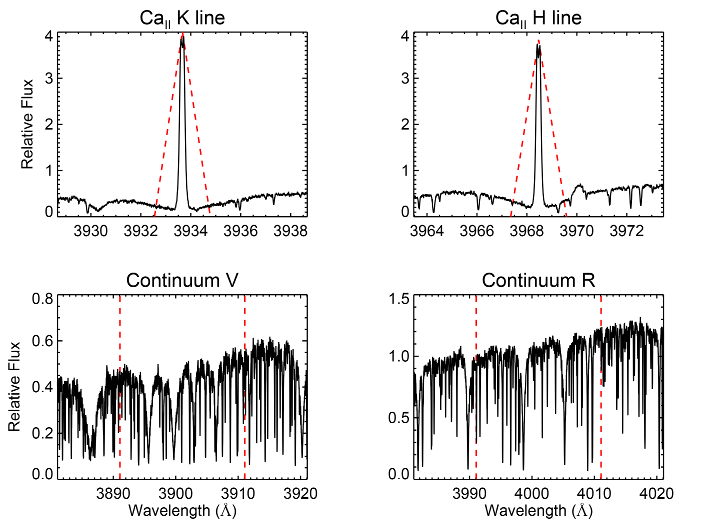}
	\caption{CA II H\&K filter of the spectrum of star GJ176, with the same shape as the Mount Wilson Ca II H\&K passband.}
	\label{ca_ii_hk_filter}
\end{figure}

Following the original Mount Wilson Observatory  procedure to characterize stellar activity,  we define the same spectral passband as used in the original Mount Wilson HKP-2 spectrophotometer   \citep{Vaughan1981} and  compute the HARPS S-index. For the CA II H\&K line cores, we define two triangular-shaped passbands with an FWHM of 1.09~{\AA} centred at 3968.470~{\AA} and 3933.664~{\AA} respectively. For the continuum bands we define two 20~{\AA} wide bands centred at 3901.070~{\AA} (V) and 4001.070~{\AA}(R). Figure~\ref{ca_ii_hk_filter} shows the shape of the filters. 

Then the S-index is defined as: 

\begin{equation}
   S=\alpha {{N_{H}+N_{K}}\over{N_{R}+N_{V}}},
\end{equation}
\noindent where $N_{H},N_{K},N_{R}$ and $N_{V}$ represent the total flux in each passband,  while $\alpha$ is a calibration constant usually fixed  at the value of 2.3. As explained in \citet{Lovis2011}, the Mount Wilson spectrophotometer is designed so that the H \& K bands get 8 times as much flux as in a normal spectrograph, so in order to be at the same scale we fixed the calibration constant as 2.3 times  8. 

\begin{figure}
	\includegraphics[width=\linewidth]{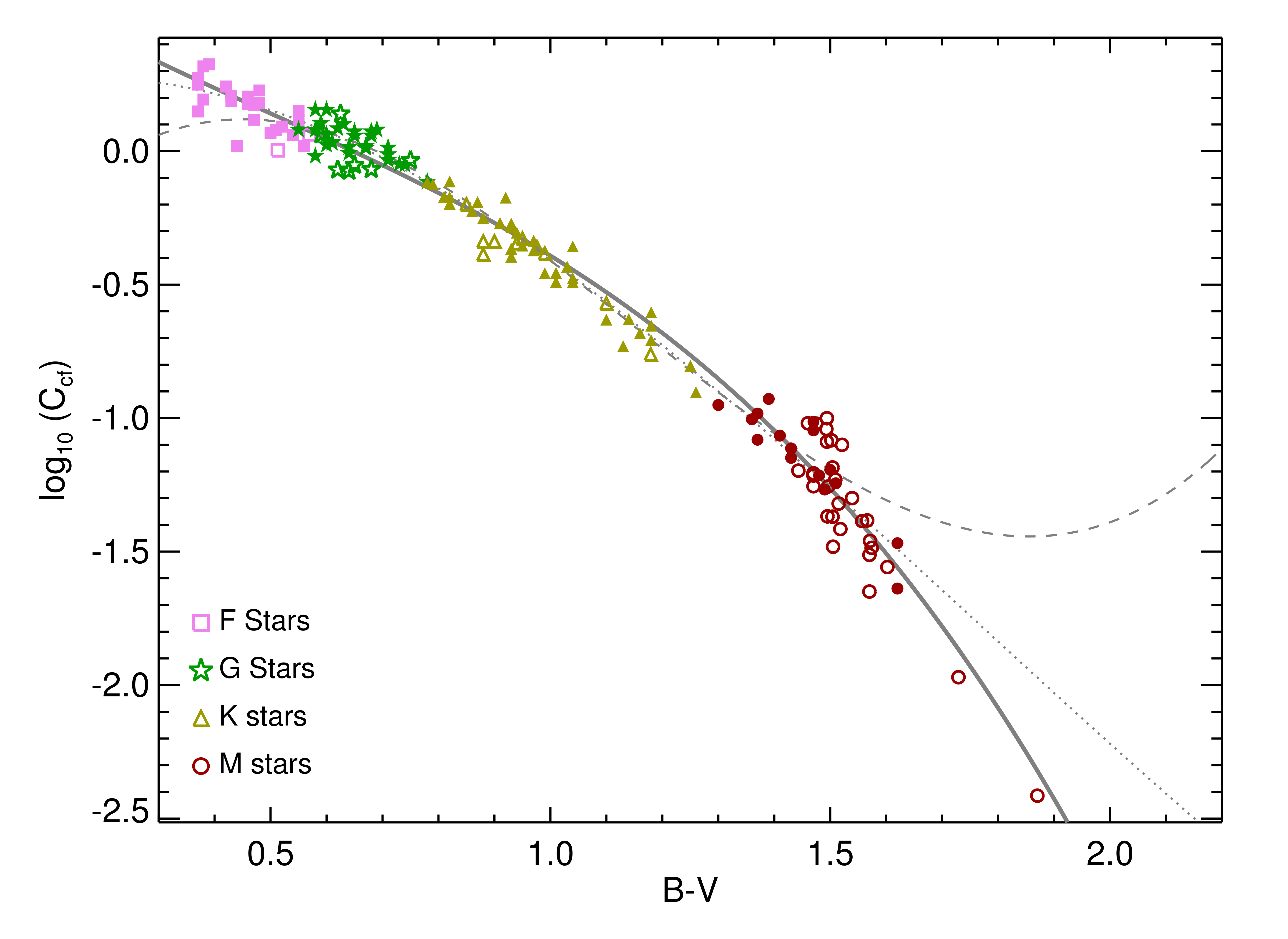}
	\caption{Conversion factor C$_{\rm cf}$ against $B-V$ colour for the combination of our sample stars (empty symbols) and the  \citet{Rutten1984} sample  (filled symbols).The curves are third-order least-squares polynomial fits. The solid line represents our relation, the dotted line represents the \citet{Rutten1984} relation and the dashed lines represents the \citet{Middelkoop1982} relation.}
	\label{CF}
\end{figure}

The S index gives the Ca II H\&K core flux normalized to the neighbouring continuum. We use this index to study periodicities related to the rotation of the star. This index contains both photospheric and chromospheric contributions. This does not have any impact when analysing one star, but to be able to compare different stars we have to subtract the photospheric contribution and normalize the chromospheric flux to the bolometric luminosity of the star. Following \citet{Noyes1984}, we compute the $R'_\textrm{HK}$, which is defined as: 

\begin{equation}
   R'_\textrm{HK}=1.34 \cdot 10^{-4} \cdot C_{\rm cf}(B-V) \cdot S-R_{\rm phot}(B-V),
\end{equation}

\noindent where $C_{\rm cf}(B-V)$ is the conversion factor that corrects flux variations in the continuum passbands and normalizes to the bolometric luminosity. The \citet{Noyes1984} original bolometric correction is based on the  \citet{Middelkoop1982} calibration, which is still widely used nowadays. But this calibration is only reliable in a narrow colour range (from $B-V \sim 0.45$ to $\sim 1.2$). Our sample also contains stars outside that range. This bolometric correction has been extended at least once \citep{Rutten1984}, but our sample still exceeds the range of application of the updated calibration.  In our case, we can directly measure the bolometric correction from the spectra. To do so we followed the original \cite{Middelkoop1982} and  \cite{Rutten1984} procedure. The factor $C_{\rm cf}$ is defined as as: 

\begin{equation}
   C_{\rm cf}=(S_{R}+S_{V}) \cdot 10^{-4} \cdot 10^{0.4 (m_v + BC)},
\end{equation}

\noindent where $S_{R}$ and $S_{V}$ are the mean fluxes measured in the continuum passbands, $m_V$ is the visual magnitude in the $V$ band and $BC$ is the bolometric correction according to \citet{Johnson1966}. Figure~\ref{CF} shows how our measured values of the bolometric correction compare to the $B-V$  colour  of the stars. We included the measurements from \citet{Rutten1984}  for comparison. 

For our stars we are using the measured bolometric correction to the index instead of the calibration, but for the future we have decided to  provide a new calibration based on our own measurements and those by \citet{Rutten1984}. Fitting a third-order least-squares polynomial to the data, we obtain the following correction factor, which is valid in the range from $B-V \sim 0.4$ to $B-V \sim 1.9$.

\begin{equation}
\begin{split}
  \log_{10}C_{\rm cf}=0.668(B-V)^{3}-1.270(B-V)^{2}\\-0.645 (B-V)-0.443 .
\end{split}
\end{equation}

 Figure~\ref{ca_ii_hk_filter} shows the shape of the H \& K bands, they are wide enough to ensure that we do not miss light from the core emission line. But this also  allows  some photospheric contribution to the measured flux in the wings of the band.  Let us call the photospheric contribution to the H \& K passbands $R_{\rm phot}$. This contribution was modelled by \citet{Hartmann1984}, who gives a useful relation for the cases where there is no direct access to spectroscopic data, but in our case we can  measure this parameter. The $R_{\rm phot}$ is just: 

\begin{equation}
   R_{\rm phot}=1.34 \cdot 10^{-4} \cdot C_{cf}(B-V) \cdot S_{\rm phot}  .
\end{equation}

So, combining this with Eq.\ 3, we get:

\begin{equation}
   R'_\textrm{HK}=1.34 \cdot 10^{-4} \cdot C_{\rm cf}(B-V) \cdot (S - S_{\rm phot}) .
\end{equation}

We measured the $S_{\rm phot}$ using Eq.\ 1 again. Following \citet{Hartmann1984}, we subtract the core emission line from the H \& K passbands in the spectrum where we recorded the lowest activity, measure the residual flux inside the bands and then compute the index. To do that we defined the limits of the core emission line inside the filter, as a 0.7~{\AA} rectangular window around the core-emission lines for F-G-K stars and a 0.4~{\AA} window for M-dwarfs, and measured the flux outside those limits. The best fit to the data is given by the following equation, but we note  that a larger stellar sample would be convenient  to  establish a proper calibration:

\begin{equation}
\begin{split}
   \log_{10}(R_{\rm phot})= 1.48 \cdot 10^{-4} \cdot exp[-4.3658 \cdot (B-V)]
\end{split}
\end{equation}

\begin{figure}
	\includegraphics[width=\linewidth]{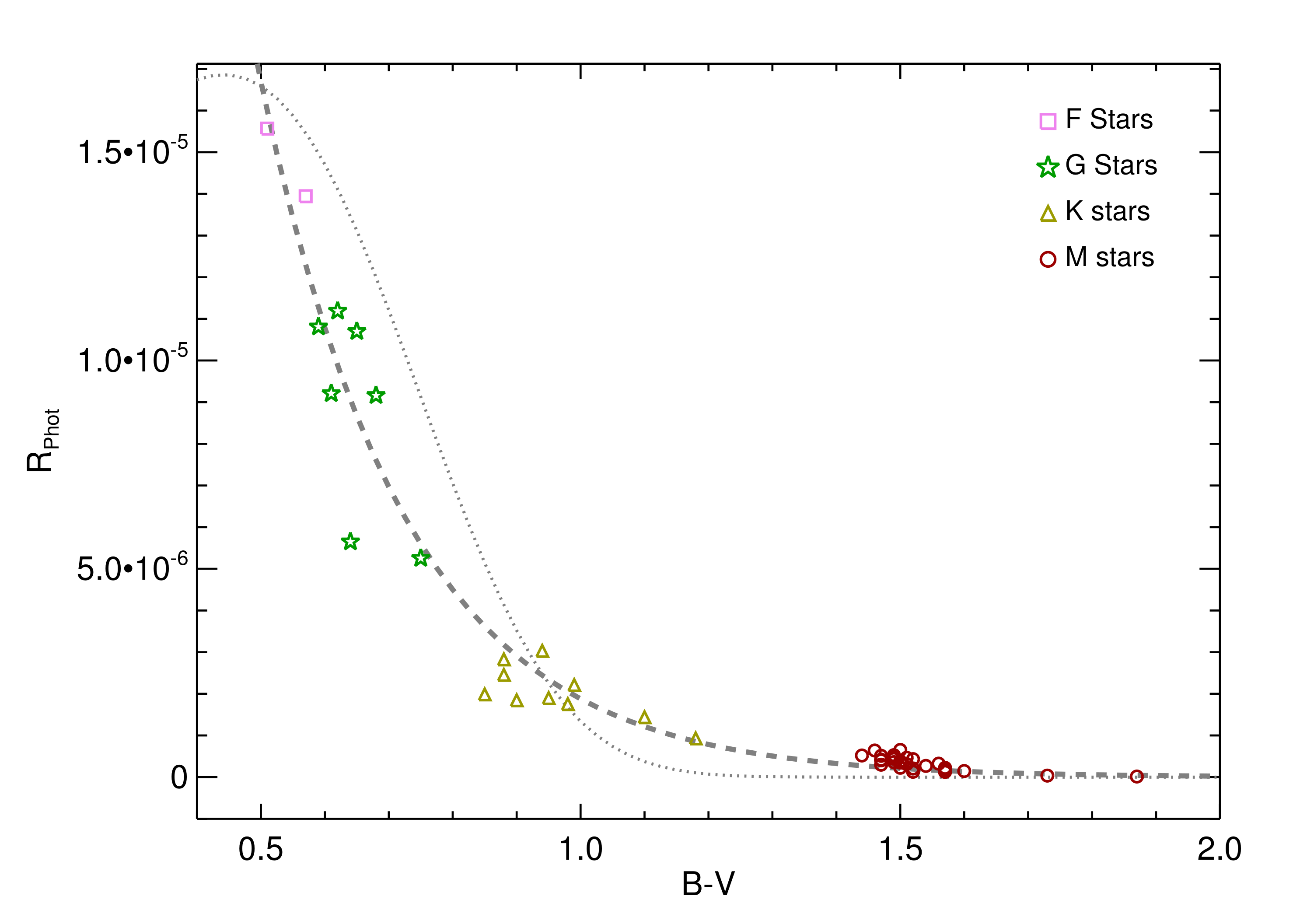}
	\caption{Measured $R_{\rm phot}$ against the colour of our sample stars. The grey dashed line shows the best fit to the data. The grey dotted line shows the original calibration by \citet{Hartmann1984}. Differences between these curves could  be due to a slightly different behaviour of the instruments on which they are based.}
	\label{Rphot}
\end{figure} 

The quantity we are finally using to compare the chromospheric acitivty level of  different stars is  $\log_{10}(R'_\textrm{HK})$. Using our direct measurements for each of the quantities involved in the definition of the index ensures the validity of the index for the full spectral range of our sample. Our new calibration provides an extension of the $\log_{10}(R'_\textrm{HK})$ index beyond the spectral range for which it was originally designed.

\begin{figure}
	\includegraphics[width=\linewidth]{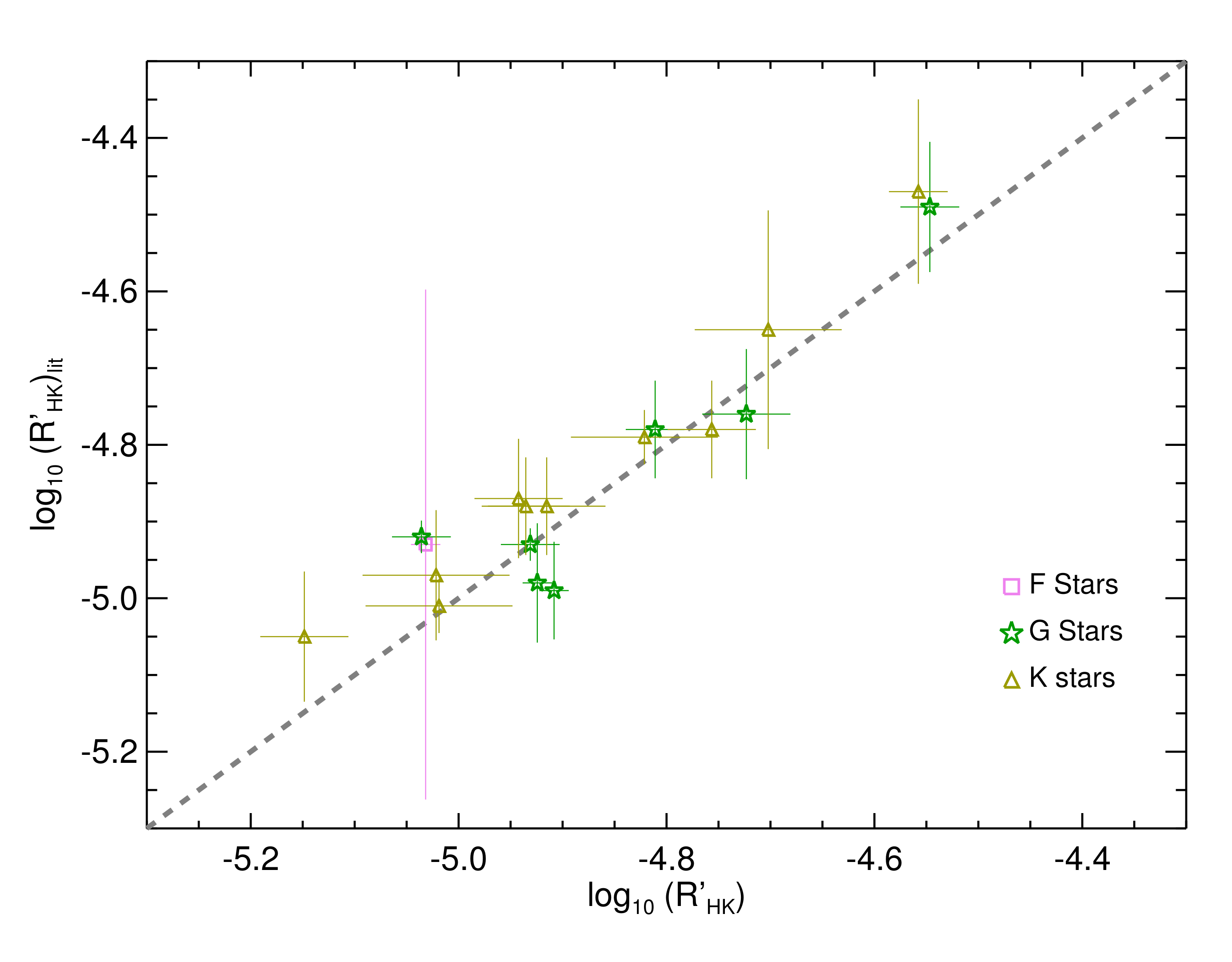}
	\caption{Comparison between the $\log_{10}(R'_\textrm{HK})$ values found in the literature and our measured $\log_{10}(R'_\textrm{HK})$. The dashed grey line represents the 1:1 line.}
	\label{RHK_RHK}
\end{figure}

Figure~\ref{RHK_RHK} and Table~\ref{tab:RHK_Parameters} shows how our $\log_{10}(R'_\textrm{HK})$ values compare to  previous measurements in the  literature for several stars in our sample. Table~\ref{tab:Data_Stars} lists the $S$, $S_{\rm phot}$ and $S_{R}+S_{V}$ values for each star. 

\begin {table}
\begin{center}
\caption {$\log_{10}(R'_\textrm{HK})$ comparison data \label{tab:RHK_Parameters}}
    \begin{tabular}{ l l l  l l l} \hline
   Star & Sp.\ Type & $\log_{10}(R'_\textrm{HK})_{\rm Lit}$& $\log_{10}(R'_\textrm{HK})$ &  Ref. \\ \hline
   HD30495	&	G2	&	--4.49	$\pm$ 0.12 &	--4.52  $\pm$  0.03 & 2,4\\   
   HD224789	&	K1	&	--4.47  $\pm$ 0.17 &	--4.55  $\pm$  0.03 &  1,2\\
   Corot-7	&	K3	&	--4.65	$\pm$ 0.22 &--4.70  $\pm$  0.07 &  3\\
   HD63765	&	G9	&	--4.76	$\pm$ 0.12 &	--4.77  $\pm$  0.04 &  1,2,5\\
   HD104067	&	K3	&	--4.78	$\pm$ 0.09 &	--4.77  $\pm$  0.04 &  2,5\\
   HD1320	&	G2	&	--4.78	$\pm$ 0.09 &--4.79  $\pm$  0.03 &  2\\  
   HD125595	&	K4	&	--4.79	$\pm$ 0.05 &	--4.82  $\pm$  0.07 &  5\\
   HD134060	&	G2	&	--4.98	$\pm$ 0.11 &	--4.89  $\pm$  0.01 &  1,2\\
   HD1388	&	G2	&	--4.99	$\pm$ 0.09 &	--4.89  $\pm$  0.01 & 1,2\\
   HD2071	&	G2	&	--4.93	$\pm$ 0.03 &	--4.89  $\pm$  0.03 & 1,2\\
   HD176986	&	K3	&	--4.88	$\pm$ 0.09 &	--4.90  $\pm$  0.04 &  1,2\\
   HD41248	&	G2	&	--4.92	$\pm$ 0.03 &	--4.90  $\pm$  0.03 &  6\\
   HD215152	&	K2	&	--4.88	$\pm$ 0.09 &	--4.92  $\pm$  0.06 &  1,2\\
   HD4628	&	K2	&	--4.87	$\pm$ 0.11 &	--4.94  $\pm$  0.04 & 2,4\\ 
   HD40307	&	K2	&	--5.01	$\pm$ 0.05 &	--5.02  $\pm$  0.07 &  1,2\\
   HD85512	&	K6	&	--4.97	$\pm$ 0.12 &	--5.02  $\pm$  0.07 &  1,2,7\\
   HD1581	&	F9	&	--4.93	$\pm$ 0.47 &	--5.03  $\pm$  0.01 &  1,2\\
   HD77338	&	K2	&	--5.05	$\pm$ 0.12 &	--5.11  $\pm$  0.04 &  2\\
    \hline
\label{fit_parameters}
\end{tabular}  
\end{center}
\begin{flushleft}
Literature values are the mean  of the values reported in different sources  and the error bars are  the standard deviations of the different measurements for a given star. Uncertainties in our reported values are the standard deviation of the individual measurements.  \\
\textbf{References:} 1 - \citet{Lovis2011},
2 - \citet{Pace2013}, 
3 - \citet{Queloz2009},
4 - \citet{Noyes1984}, 
5 - \citet{Segransan2011},
6 - \citet{Jenkins2013}, 
7 - \citet{Pepe2011}
\end{flushleft}
\end {table}

\subsection{The HARPS H$\alpha$ index}

\begin{figure}
	\includegraphics[width=\linewidth]{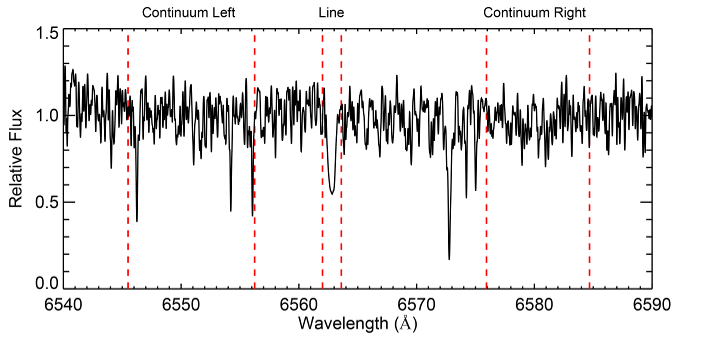}
	\caption{Spectrum of the M-type star GJ176 showing the H$\alpha$ filter passband and continuum bands.}
	\label{halpha_filter}
\end{figure}

In the case of the H$\alpha$ index we use a simpler passband following \cite{GomesdaSilva2011}. It consists of a rectangular bandpass with a width of 1.6~{\AA} and centred at 6562.808~{\AA} (core), and two continuum bands of 10.75~{\AA} and a 8.75~{\AA} wide centred at 6550.87~{\AA} (L) and 6580.31~{\AA} (R), respectively, as seen in Figure~\ref{halpha_filter}. 

Thus, the H$\alpha$ index is defined as

\begin{equation}
   H\alpha_{\rm Index}={{H\alpha_{\rm core} }\over{H\alpha_{L} +H\alpha_{R}}}.
\end{equation}
 
\section{TIME-SERIES ANALYSIS OF ACTIVITY INDICATORS AND ROTATION PERIODS}

We perform a time-series analysis of the Ca II H\&K and H$\alpha$ activity indicators to search for periodicities related to stellar rotation. For many stars our spectroscopic time series show that the variability of the S- and the H$\alpha$ indexes is beyond what can be explained by the error bars of the measurements. We therefore decided to search for any periodic behaviour in the series of data. We computed  power spectra by fitting a sinusoidal model for each trial frequency using the MPFIT routine  \citep{Markwardt2009} and then  build a Lomb--Scargle periodogram \citep{Lomb1976, Scargle1982}, following \citet{Cumming2004} and taking into account the individual error bars for each point, which makes it equivalent to the Generalised Lomb--Scargle Periodogram \citep{Zechmeister2009}.

\begin{figure}
	\includegraphics[width=\linewidth]{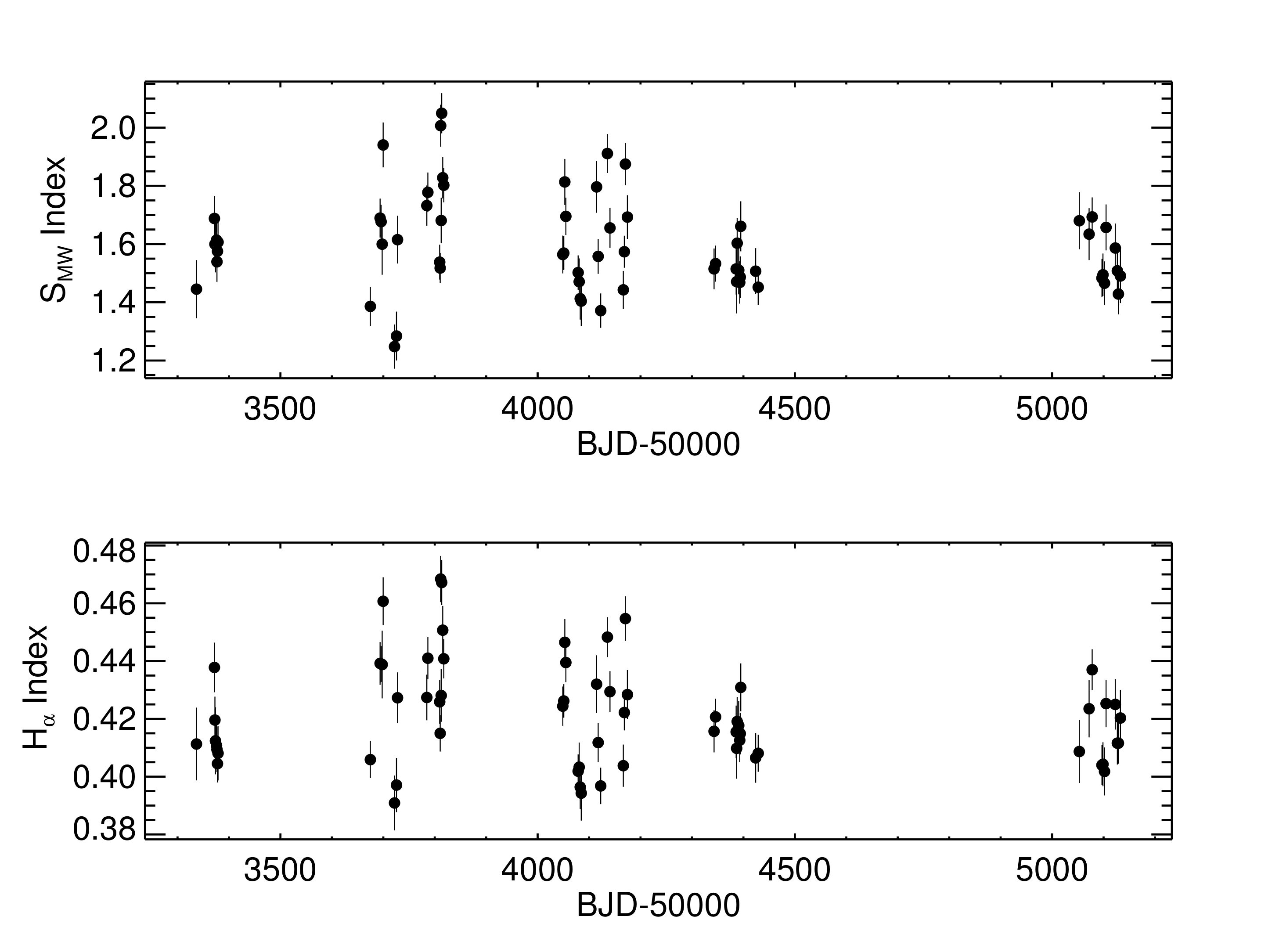}
	\caption{S$_{\rm MW}$ (top) and H$\alpha$ (bottom) index time series for the star GJ176.}
		\label{timeseries}
\end{figure}

\begin{figure}
	\includegraphics[width=\linewidth]{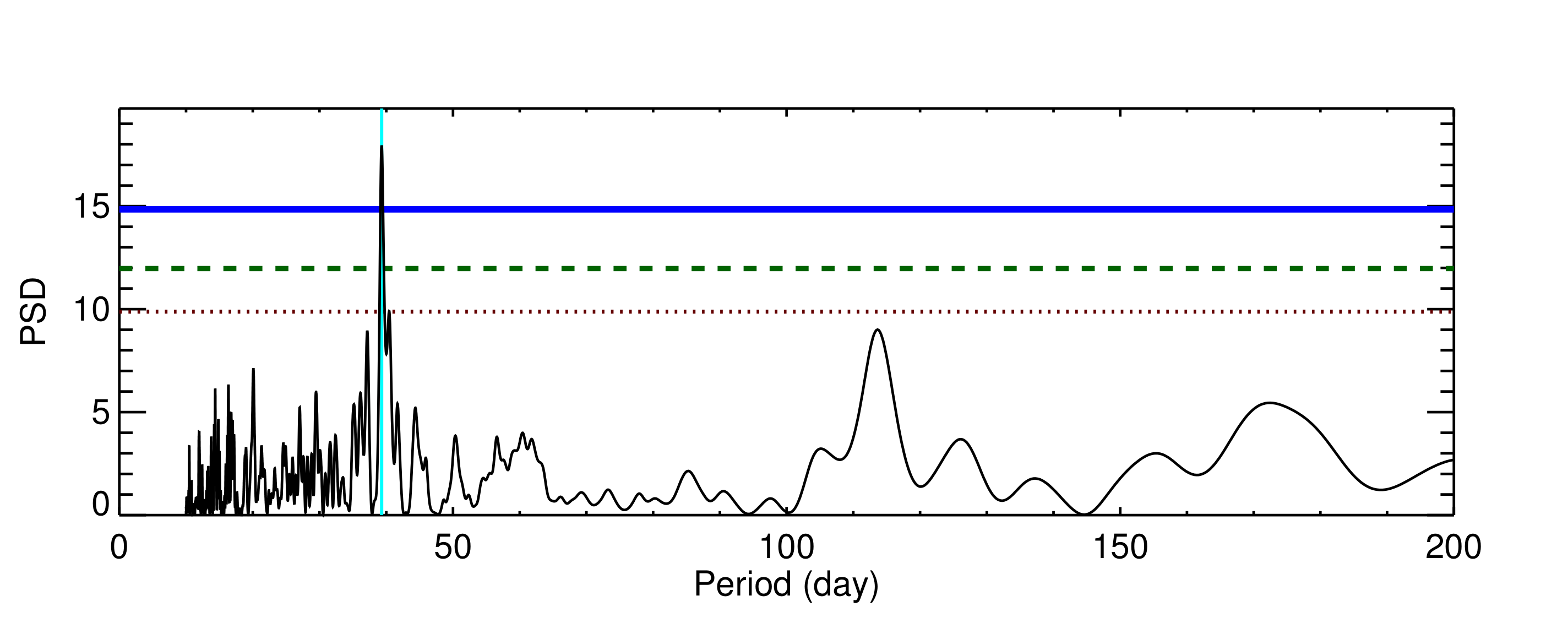}
	\includegraphics[width=\linewidth]{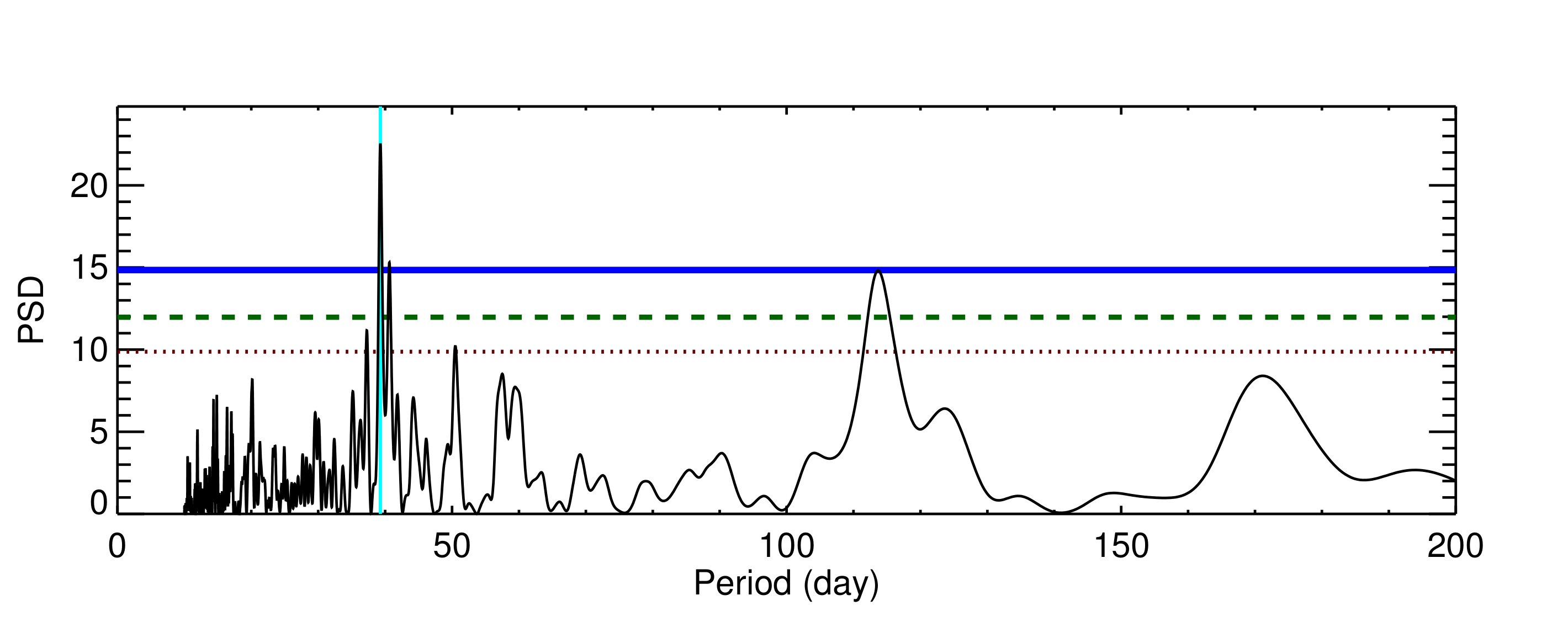}
	\caption{S$_{\rm MW}$ (top) and H$\alpha$ (bottom) index power spectrum showing a peak at $\sim$39.3 days for GJ176. Red dotted line represents the 1$\sigma$ detection threshold, green dashed line the 2$\sigma$ threshold and blue solid line the 3$\sigma$ detection threshold.}
	\label{periodogram}
\end{figure}

\begin{figure}
	\includegraphics[width=\linewidth]{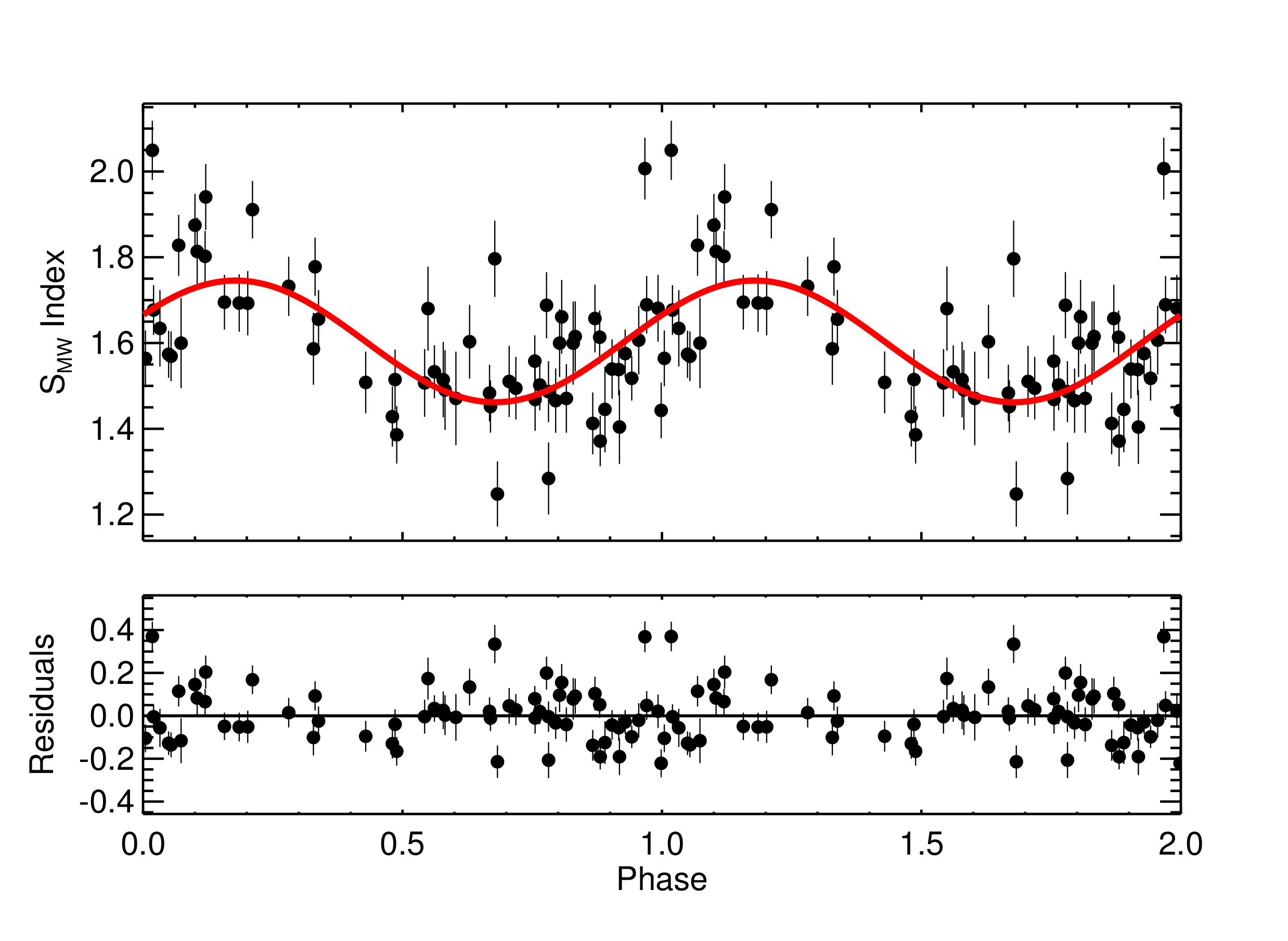}
	\includegraphics[width=\linewidth]{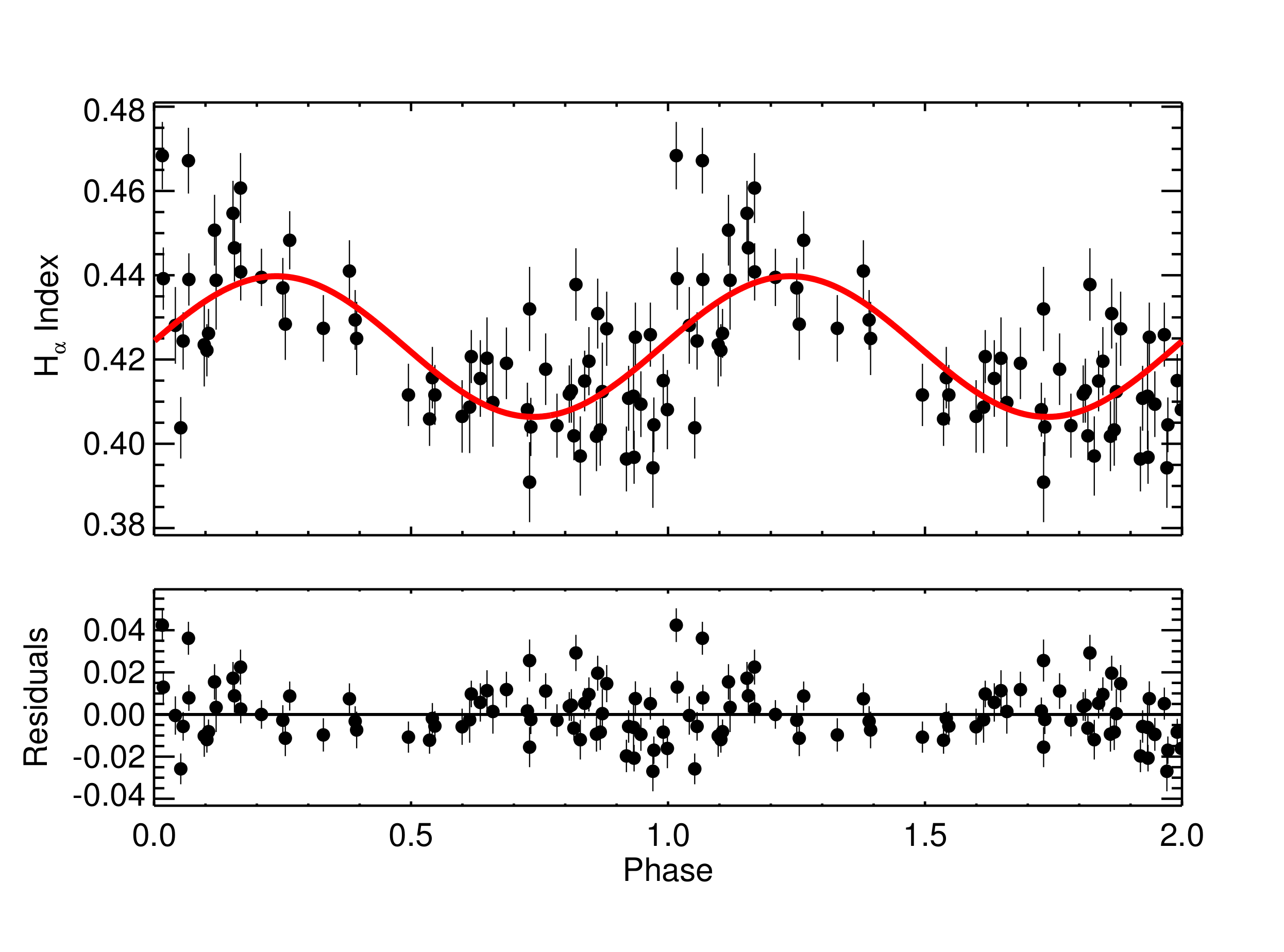}
	\caption{Fit of the S$_{\rm MW}$ (top) and H$\alpha$ (bottom) index data using the 39.3 days period given by the power spectrum for the star GJ176.}
	\label{FIT}
\end{figure}

To evaluate the false alarm probability of any peak in the periodogram, we follow \citet{Cumming2004}, who modified the work by \citet{HorneBaliunas1986} to obtain the spectral density thresholds for a desired false alarm level. This means our false alarm probability is defined as $FAP = 1 - [1 - P (z > z_{0}]^{M}$ where $P (z > z_{0}) = exp (-z_{0})$ is the  probability of $z$ being greater than $z_{0}$, with $z$ the target spectral density, $z_{0}$ the measured spectral density and $M$ the number of independent frequencies. We search for the power values corresponding to 31.7$\%$, 4.6$\%$ and 0.3$\%$ false alarm probability, equivalent to 1$\sigma$, 2$\sigma$ and 3$\sigma$ detections. After building the power spectrum, we use the best frequency found as an initial estimate for a new sinusoidal fit of the time series. 

\begin{figure*}
\includegraphics[width=\textwidth]{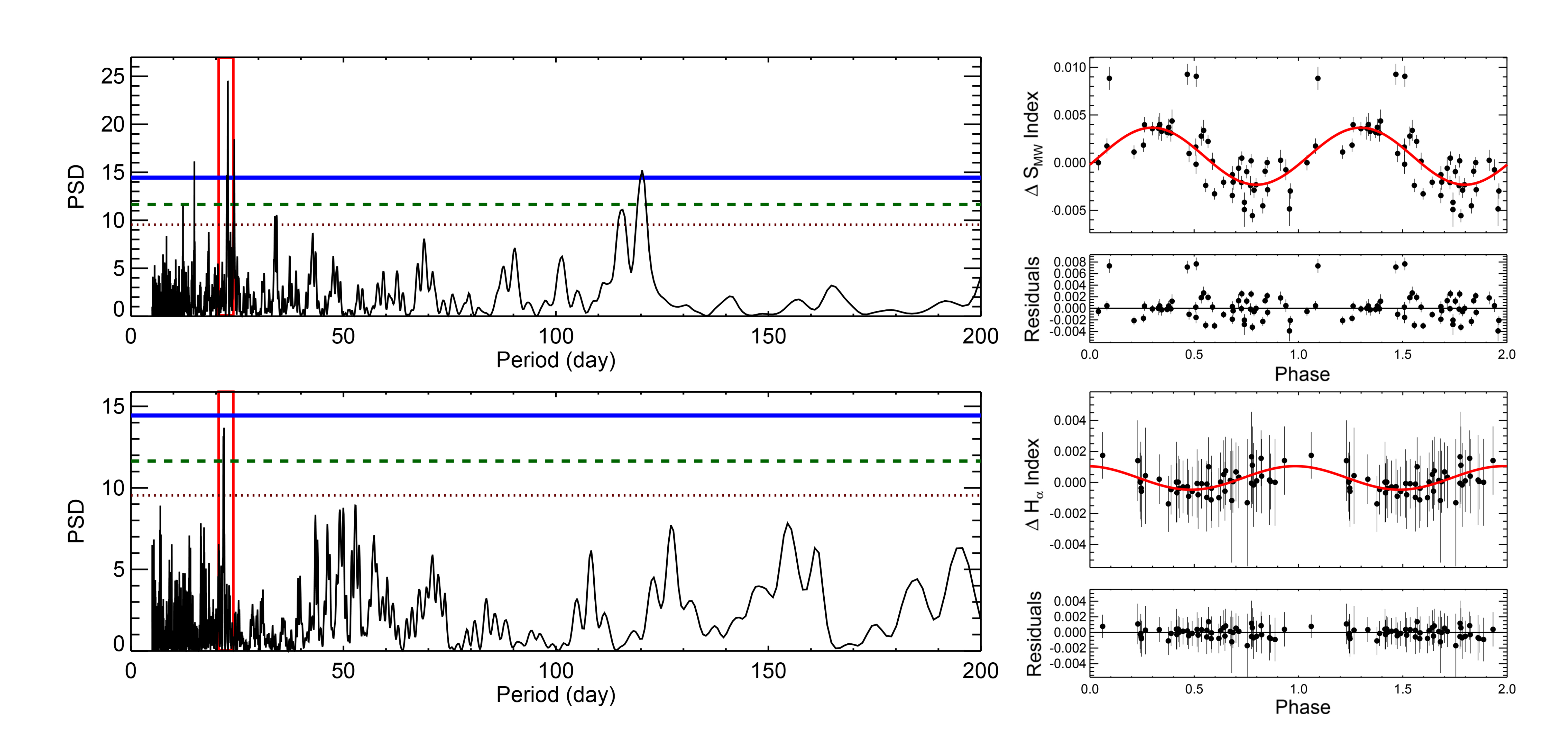}
\caption{S$_{\rm MW}$ (top) and H$\alpha$ (bottom) index power spectrum and fit for the solar type star HD2071. Red dotted line represents the 1$\sigma$ detection threshold, green dashed line the 2$\sigma$ threshold and blue solid line the 3$\sigma$ detection threshold.}
\label{HD2071}
\end{figure*}

\begin{figure*}
\includegraphics[width=\textwidth]{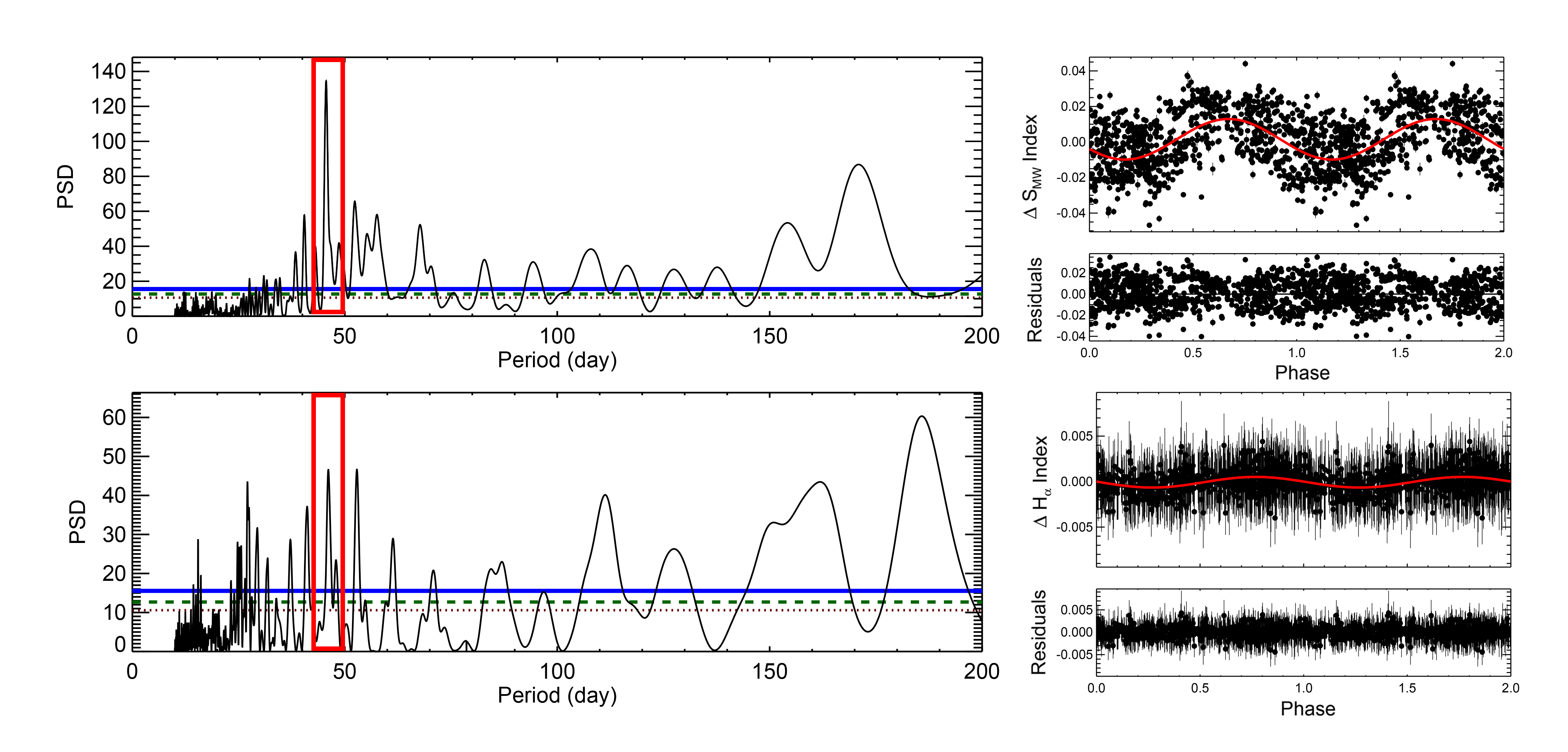}
\caption{S$_{\rm MW}$ (top) and H$\alpha$ (bottom) index power spectrum and fit for the K-type star HD85512. Red dotted line represents the 1$\sigma$ detection threshold, green dashed line the 2$\sigma$ threshold and blue solid line the 3$\sigma$ detection threshold.}
\label{HD85512}
\end{figure*}

We have not taken into account either very  fast rotators or the long-term magnetic cycles of the stars. The study of these very short and very long period modulations is beyond the  scope of this paper. Thus, the typical rotation periods we search for  range from 5 to 200 days. These limits were defined after studying the typical rotation period of late type low activity main sequence stars. To cope with the presence of long term cycles we proceed to fit them and remove them before performing the search for periodicities. 

To illustrate the procedure, we display in Figs~\ref{timeseries},~\ref{periodogram} \&~\ref{FIT} the results for the M3 dwarf star GJ176, for which a rotational period of 38.92 $\pm$ 3.9 days has been claimed by \citep{Kiraga2007}, and for which an in-depth study has been carried out by \citet{Robertson2015}. Our spectroscopic periods obtained from Ca II H\&K and H$\alpha$ are consistent to  within 0.1 days, and the average value is 39.3 days (see Table~\ref{tab:RotationPeriods}), which  is more precise and in good agreement with the previous value.

In the example of GJ176 the detection of the rotation period is straightforward (see Fig.~\ref{periodogram}). Both indexes give the same result and the periodogram is clear. But that is not always the case. In a few cases the two indicators do not provide such a good agreement. Typically this happens when we have a weak detection in one of the indicators (in $H\alpha$ in early type stars, Ca II H\&K in some late Ms). As the determined period is a weighted average between the measurements of both indicators, using the false alarm probability as weight, the strong detection becomes dominant for these stars. Sometimes multiple peaks are present because of inappropriate sampling and one could be in the situation that one of the indexes shows a harmonic of the rotation period with a stronger spectral density than the actual rotation period \citep{Boisse2011}. In such circumstances we rely in the cleanest periodogram as a starting point for the analysis. Figure~\ref{HD2071} and~\ref{HD85512} illustrate cases where the determination of the rotation period is not straightforward, one for a solar type star, another one for a K-type star.

Figure~\ref{HD2071} shows the analysis of the solar type star HD2071 after correcting for a long term magnetic cycle. Ca II H$\&$K gives us a significant detection at $\sim$22 days with a good fit to the data. Then the  H$\alpha$ index, despite the low amplitude of the curve, gives us mostly the same peak. Significance is lower because of the low amplitude, but still around 1\% of false alarm probability. Having both results we can conclude a rotation period for this star of 22.8$\pm$0.7 days.

Figure~\ref{HD85512} shows the analysis of the K-type star HD85512 after correcting for a long term cycle. In this situation we have a clear detection in the Ca II H$\&$K index and a good fit to the data for a period of $\sim$45 days. In the H$\alpha$ index the detection is not as clear, we have several peaks because of the low amplitude of the signal and window effects. The $\sim$180 days peak is a window effect. Then we have three strong peaks at $\sim$25, $\sim$46 and $\sim$52 days respectively, being the $\sim$46 days the strongest and more consistent with the peak in the Ca II H$\&$K index. Then the we weighted average the two measurements and obtain 45.9$\pm$0.4 days as our best measurement. 

In Table~\ref{tab:RotationPeriods} labels explaining the different cases are given when needed.

\subsection{Rotation Periods}

\begin{figure}
	\includegraphics[width=\linewidth]{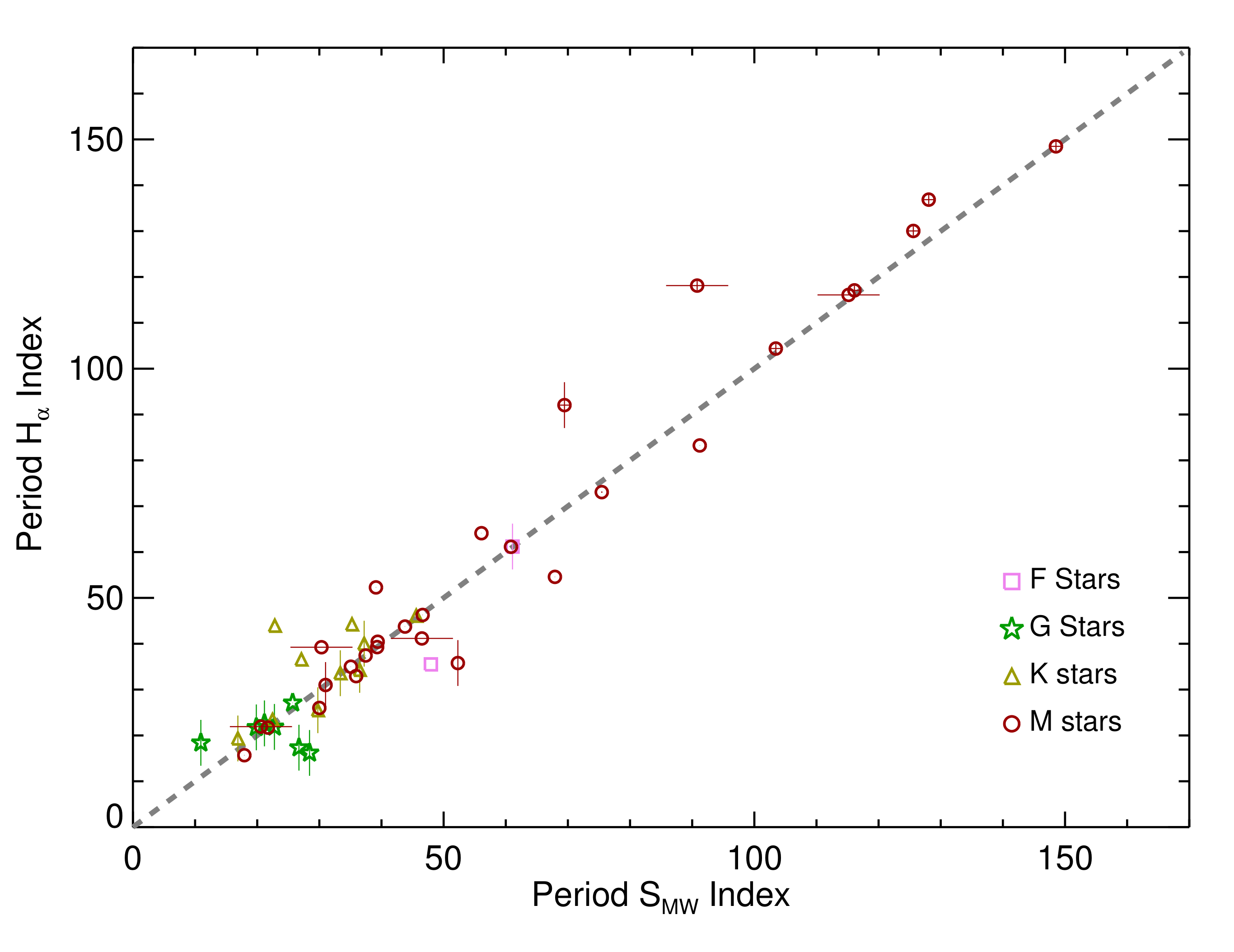}
	\caption{Comparison of the spectroscopic periods measured from the  Ca$_{\rm HK}$ and H$\alpha$ indexes. The line shows the 1:1 relation.}
	\label{period_ca_ha}
\end{figure}

Table~\ref{tab:RotationPeriods} provides the rotation periods for the stars for which we have been able to detect a periodic signal. Final periods were measured by averaging the detections in the Ca$_{\rm HK}$ and H$\alpha$ indexes, weighting each one with the spectral density of the peaks in the corresponding periodigrams. The listed uncertainties were adopted conservatively, as the difference between the periods obtained from the two indexes detections. Fig.~\ref{period_ca_ha} depicts the agreement between the periods derived from the  Ca$_{\rm HK}$ and H$\alpha$.

\begin{figure}
	\includegraphics[width=\linewidth]{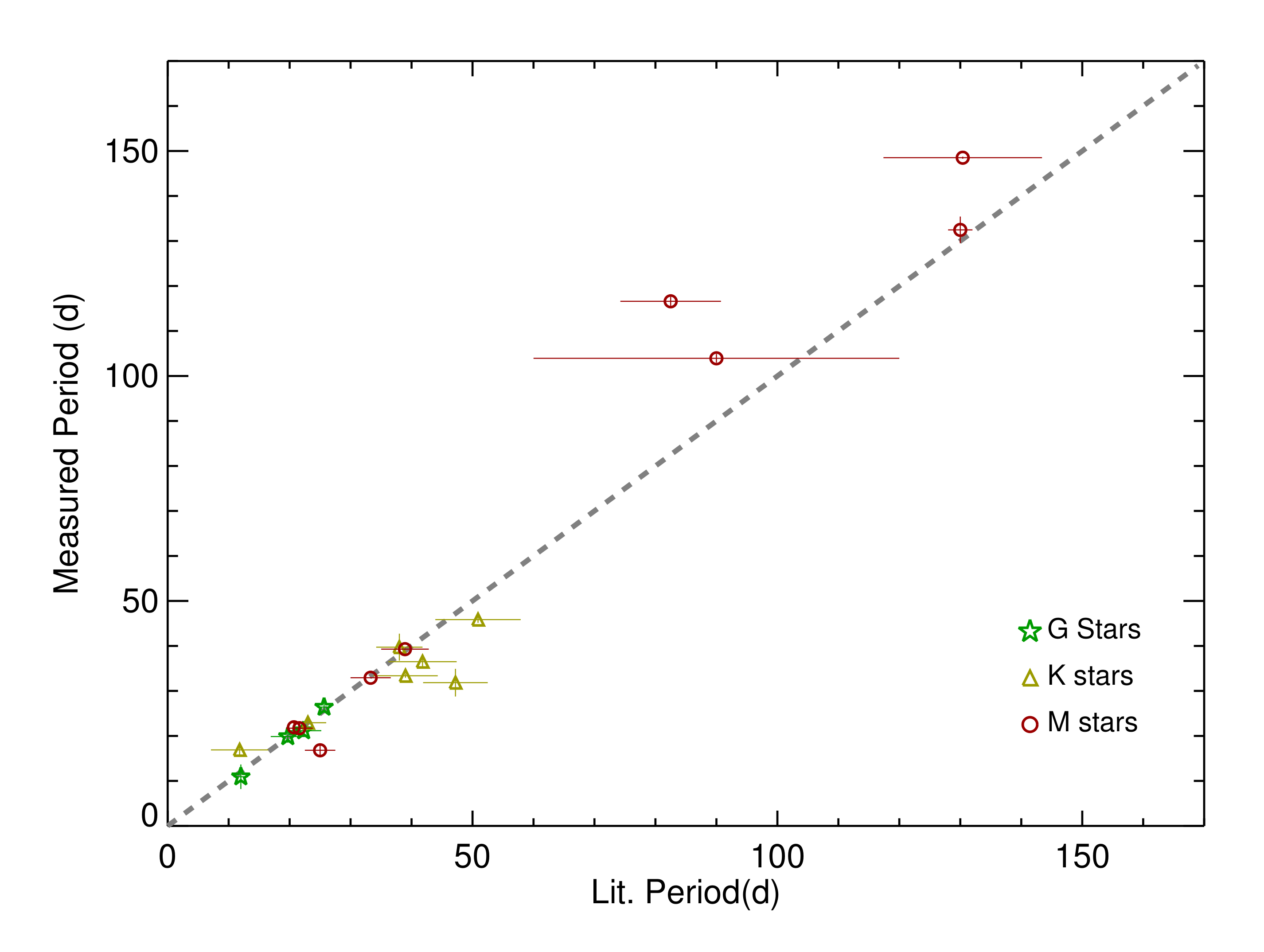}
	\caption{Proposed spectroscopic periods versus previous period determinations available in the literature. Only a handful of stars in the sample have previous determinations. The line shows the 1:1 relation.}
	\label{period_lit}
\end{figure}

Figure~\ref{period_lit} and Table~\ref{tab:RotationPeriods} show how our detections compare with previous determinations of the rotation periods of stars in our sample. There is, in general, very good agreement.   

\begin{figure*}
\includegraphics[width=\textwidth]{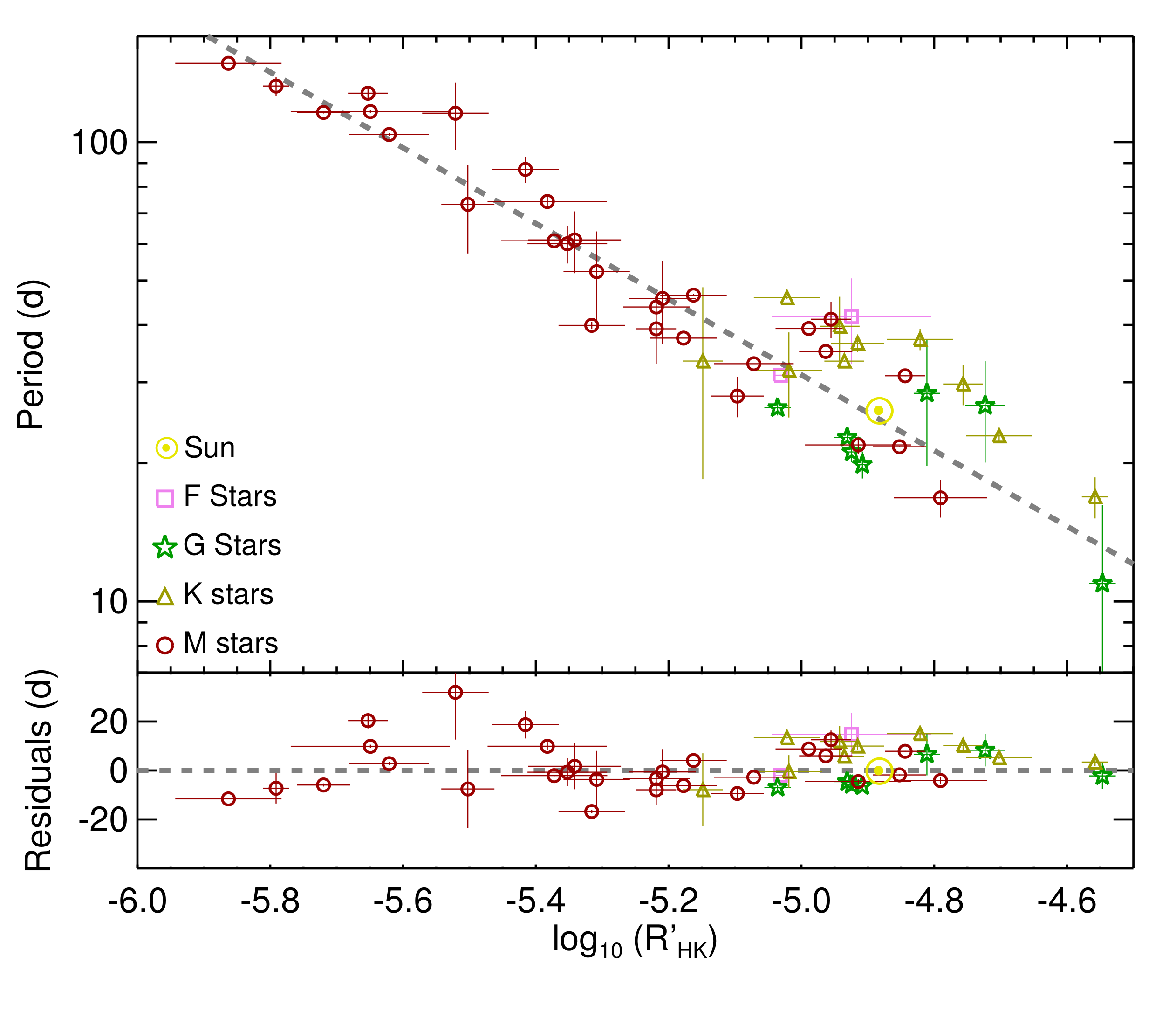}
\caption{Rotation period versus chromospheric activity level $\log_{10}(R'_\textrm{HK})$. The dashed line shows the global fit to the full dataset. Error bars in rotation periods reflect differences in the period measurements obtained from   the two activity indexes.  Error bars in the $\log_{10}(R'_\textrm{HK})$ come from the standard deviation of individual measurements. For the Sun we adopted the rotation period of 26.09 days \citep{Donahue1996} and a $\log_{10}(R'_\textrm{HK})$ of $-4.906$ \citep{Mamajek2008}}
\label{period_rhk}
\end{figure*}

Figure~\ref{period_rhk} depicts the relation between our spectroscopic periods and the mean activity level measured using the Ca II H\&K lines,$\log_{10}(R'_\textrm{HK})$. The two quantities show an evident correlation that is  described well by the following linear fit:  

\begin{equation}
   \log_{10}(P)= a \cdot \log_{10}(R'_{\rm HK}) + b,
\end{equation}

\noindent with  values for the parameters $ a$ and $b$ as  given  in Table~\ref{tab:Parameters}, being the scatter of the residuals after the fit smaller than the 17\% of the measured periods.

This relationship appears to be very reliable for M-type dwarfs, which populate the full range of stellar activity and rotation periods in the sample. We think this relation could be used to estimate the rotation periods of M-dwarfs  from the more simple measurement of the activity level.   While   F-, G-, and K-type stars also  lie on the relation, the lack of  stars of these types  at very low activity level  ($\log_{10}(R'_\textrm{HK}$) below --5 ) leads us to recommend caution  when estimating rotation periods of early-type stars via this relationship.

\begin {table}
\begin{center}
\caption {Parameters for equation 9\label{tab:Parameters}}
    \begin{tabular}{ l  l  l } \hline
Dataset & a & b  \\ \hline
Full & --0.808 $\pm$ 0.012 & --2.536 $\pm$ 0.009\\
M-type Stars & --0.753 $\pm$ 0.056 & --2.219 $\pm$ 0.389\\
$-0.5 < {\rm [Fe/H]} < -0.1$ & --0.821 $\pm$ 0.025 & --2.610 $\pm$ 0.066\\
Solar [Fe/H] & --0.773 $\pm$ 0.017 & --2.347 $\pm$ 0.002\\
$+0.1 < {\rm [Fe/H]} < +0.3$ & --1.063 $\pm$ 0.147 & --3.817 $\pm$ 0.640\\
\label{fit_parameters}
\end{tabular}  
\end{center}
\end {table}

\begin {table*}
\begin{center}
\caption { Rotational Periods and Chromospheric Activity\label{tab:RotationPeriods}}
    \begin{tabular}{ l  l  l  l  l l l l l l} \hline
Name & SpTp & B-V  &  $\log_{10}(R'_\textrm{HK})$~$^{a}$ & Period (d)~$^{b}$ & Sig.~$^{c}$ & Lit. Period (d) & Reference & Comments\\ \hline
HD30495 & G2 & 0.64 & --4.52  $\pm$  0.03	&	10.9  $\pm$  5.3 & $3\sigma$ & 12.0 $\pm$ 1.2$^{d}$ & \citet{Wright2004} & 1,3\\
HD224789 & K1 & 0.88 & --4.55  $\pm$  0.03	&	16.9  $\pm$  1.8 & $3\sigma$ & 11.8	$\pm$ 4.7 & \citet{Lovis2011} & 3\\
GJ358 & M2 & 1.52 & --4.69  $\pm$  0.10	&	16.8  $\pm$  1.6 & $2\sigma$ & 25.0 $\pm$ 2.5$^{d}$ & \citet{Kiraga2007}\\
Corot-7 & K0 & 0.85 & --4.70  $\pm$  0.07	&	22.9  $\pm$  0.7 & $3\sigma$ & 23.0  $\pm$ 3.0 & \citet{Leger2009} & 1\\
HD63765 & G5 & 0.75 & --4.77  $\pm$  0.04	&	26.7  $\pm$  6.7 & $3\sigma$ & & & 4\\
HD104067 & K3 & 0.98 & --4.77  $\pm$  0.04	&	29.8  $\pm$  3.1 & $3\sigma$ & & & 3\\
HD1320 & G2 & 0.65 & --4.79  $\pm$  0.03	&	28.4  $\pm$  8.7 & $2\sigma$ & & & 1,3\\
GJ382 & M1 & 1.48 & --4.80  $\pm$  0.06	&	21.7  $\pm$  0.1 & $3\sigma$ & 21.6  $\pm$ 2.2 & \citet{Kiraga2007}\\
HD125595 & K4 & 1.10 & --4.82  $\pm$  0.07	&	37.2  $\pm$  2.0 & $3\sigma$ & & & 1,3\\
GJ846 & M1 & 1.47 & --4.84  $\pm$  0.04	&	31.0  $\pm$  0.1 & $3\sigma$\\
HD134060 & G1 & 0.62 & --4.89  $\pm$  0.01	&	21.2  $\pm$  1.1 & $3\sigma$ & 22.3 	$\pm$ 2.9 & \citet{Lovis2011} & 1\\
HD1388 & G0 & 0.59 & --4.89  $\pm$  0.01	&	19.9  $\pm$  1.4 & $3\sigma$ & 19.7  $\pm$ 2.8 & \citet{Lovis2011} 1,3\\
HD2071 & G2 & 0.68 & --4.89  $\pm$  0.03	&	22.8  $\pm$  0.7 & $3\sigma$ & & & 1,3\\
HD41248 & G1 & 0.61 & --4.90  $\pm$  0.03	&	26.4  $\pm$  1.1 & $3\sigma$ & 25.65  $\pm$ 1.13 & \citet{Santos2014} \\
HD176986 & K3 & 0.94 & -4.90  $\pm$  0.04	&	33.4  $\pm$  0.2 & $3\sigma$  & 39.0 $\pm$ 5.3 & \citet{Lovis2011} & 3\\
Sun & G2 & &  --4.91 & & &  26.09 \\
GJ3470 & M2 & 1.50 & --4.91  $\pm$  0.11	&	21.9  $\pm$  1.0  & $3\sigma$  & 20.7  $\pm$ 0.15 & \citet{Biddle2014} & 1,2\\
HD215152 & K3 & 0.99 & -4.92  $\pm$  0.06	&	36.5  $\pm$  1.6 & $3\sigma$  & 41.8 	$\pm$ 5.6 & \citet{Lovis2011} & 1,3\\
HD25171 & F9 & 0.51 & --4.92  $\pm$  0.17	&	35.6  $\pm$  8.9 & $2\sigma$ & & & 3\\
HD4628 & K2 & 0.90 & --4.94  $\pm$  0.04	&	39.7  $\pm$  6.4 & $2\sigma$ & 38.0	$\pm$ 3.8$^{d}$ & \citet{Noyes1984} & 1,4\\
GJ205 & M1 & 1.47 & --4.96  $\pm$  0.06	&	35.0  $\pm$  0.1 & $3\sigma$ \\
GJ676A & M0 & 1.44 & --4.96  $\pm$  0.04	&	41.2  $\pm$  3.8 & $3\sigma$ & & & 2\\
GJ176 & M3 & 1.54 & --4.99  $\pm$  0.07	&	39.3  $\pm$  0.1 & $3\sigma$  & 38.92 $\pm$ 3.9$^{d}$  & \citet{Kiraga2007} \\
HD40307 & K3 & 0.95 & --5.02  $\pm$  0.07	&	31.8  $\pm$  6.7 & $3\sigma$ & 47.2 	$\pm$ 5.3 & \citet{Lovis2011} & 4\\
HD85512 & K6 & 1.18 & --5.02  $\pm$  0.07	&	45.9  $\pm$  0.4 & $3\sigma$  & 50.9 	$\pm$ 7.0 & \citet{Lovis2011} & 1\\
HD1581 & F9 & 0.57 & --5.03  $\pm$  0.01	&	31.1  $\pm$  0.1 & $2\sigma$ & & & 3\\
GJ674 & M3 & 1.57 & --5.07  $\pm$  0.08	&	32.9  $\pm$  0.1  & $3\sigma$ & 33.29 $\pm$ 3.3$^{d}$ & \citet{Kiraga2007}\\
GJ514 & M1 & 1.48 & --5.10  $\pm$  0.06	&	28.0  $\pm$  2.9 & $3\sigma$ \\
HD77338 & K2 & 0.88 & --5.11  $\pm$  0.04	&	33.4  $\pm$  10.0 & $3\sigma$ & & & 1,4\\
GJ849 & M2 & 1.50 & --5.14  $\pm$  0.04	&	39.2  $\pm$  6.3  & $3\sigma$ & & & 2 \\
GJ752A & M2 & 1.52 & --5.16  $\pm$  0.07	&	46.5  $\pm$  0.3 & $3\sigma$ \\
GJ880 & M2 & 1.51 & --5.18  $\pm$  0.07	&	37.5  $\pm$  0.1 & $3\sigma$ & & & 1\\
GJ832 & M2 & 1.50 & --5.21  $\pm$  0.07	&	45.7  $\pm$  9.3  & $3\sigma$ & & & 1,4\\
GJ536 & M1 & 1.47 & --5.22  $\pm$  0.07	&	43.8  $\pm$  0.1  & $3\sigma$ & & & 1\\
GJ526 & M1 & 1.49 & --5.31  $\pm$  0.07	&	52.3  $\pm$  1.7  & $3\sigma$ & & & 3\\
GJ436 & M1 & 1.47 & --5.32  $\pm$  0.07	&	39.9  $\pm$  0.8 & $3\sigma$ \\
GJ588 & M2 & 1.50 & --5.34  $\pm$  0.10	&	61.3  $\pm$  6.5 & $3\sigma$ & & & 1\\
GJ1 & M0 & 1.46 & --5.35  $\pm$  0.08	&	60.1  $\pm$  5.7 & $3\sigma$ & & & \\
GJ163 & M1 & 1.48 & --5.37  $\pm$  0.11	&	61.0  $\pm$  0.3  & $3\sigma$ \\
GJ357 & M3 & 1.57 & --5.38  $\pm$  0.13	&	74.3  $\pm$  1.7 & $3\sigma$ & \\
GJ876 & M3 & 1.56 & --5.42  $\pm$  0.07	&	87.3  $\pm$  5.7  & $3\sigma$ & 95.0 $\pm$ 1.0 & \citet{Nelson2015} \\
GJ433 & M2 & 1.51 & --5.50  $\pm$  0.06	&	73.2  $\pm$  16.0 & $3\sigma$ & & & 4\\
GJ273 & M3 & 1.57 & --5.52  $\pm$  0.07	&	115.6  $\pm$  19.4 & $3\sigma$ & & & 2\\
GJ667C & M3 & 1.57 & --5.62  $\pm$  0.08	&	103.9  $\pm$  0.7 & $3\sigma$ & 90.0	$\pm$ 30.0 & \citet{Robertson2014b}\\
GJ551 & M6 & 1.87 & --5.65  $\pm$  0.17	&	116.6  $\pm$  0.7  & $3\sigma$ & 82.5  $\pm$ 8.25$^{d}$ & \citet{Kiraga2007} \\
GJ701 & M2 & 1.52 & --5.65  $\pm$  0.04	&	127.8  $\pm$  3.2  & $3\sigma$ & & & 1\\
GJ877 & M2 & 1.50 & --5.72  $\pm$  0.06	&	116.1  $\pm$  0.7  & $3\sigma$ & & & 2\\
GJ581 & M4 & 1.60 & --5.79  $\pm$  0.03	&	132.5  $\pm$  6.3  & $3\sigma$ & 130   $\pm$ 2.0 & \citet{Robertson2014} \\
GJ699 & M5 & 1.73 & --5.86  $\pm$  0.11	&	148.6  $\pm$  0.1 & $3\sigma$ & 130.4 $\pm$ 13.0$^{d}$ &  \citet{Benedict1998} & 1\\ \hline
\label{table_results}
\end{tabular}  
\end{center}
\begin{flushleft}
$^{a}$Error bars in the $\log_{10}(R'_\textrm{HK})$ come from the standard deviation of individual measurements.
$^{b}$Period stands for the spectroscopic determination of rotation period in days. Uncertainties are inferred from the difference measured Ca$_{\rm HK}$ and H$\alpha$ indicators.\\
$^{c}$Significance. Related to the false alarm probability of  period detections with 3$\sigma$ equivalent to or less than 0.3\%.\\
$^{d}$For this period available in the literature, no uncertainty is provided, and thus 10\% of measured period is adopted.\\
\textbf{Notes:} \citet{Kiraga2007}, \citet{Leger2009}, \citet{Biddle2014}, \citet{Benedict1998} and \citet{Biddle2014} periods are derived using photometric time series. 
\citet{Noyes1984}  derives the periods using an activity--rotation model. 
\citet{Lovis2011} periods are derived using \citet{Mamajek2008} activity--rotation model. 
\citet{Wright2004} uses \citet{Noyes1984} procedure.
\citet{Santos2014} uses activity diagnosis analysis on the radial velocity time series. 
\citet{Robertson2014b} uses \citet{Engle2011} relation based on X-ray emission.
\citet{Nelson2015} uses $H_{\alpha}$ time-series analysis.
Sun values taken from \citet{Donahue1996, Mamajek2008}.\\ 
\textbf{Comments:} 1 - Presence of a long term magnetic cycle. 2 - Weak Ca$_{\rm HK}$ detection. 3 - Weak H$\alpha$ detection. 4 - Two different detections.  
\end{flushleft}
\end {table*}

\begin{figure*}
\includegraphics[width=\textwidth]{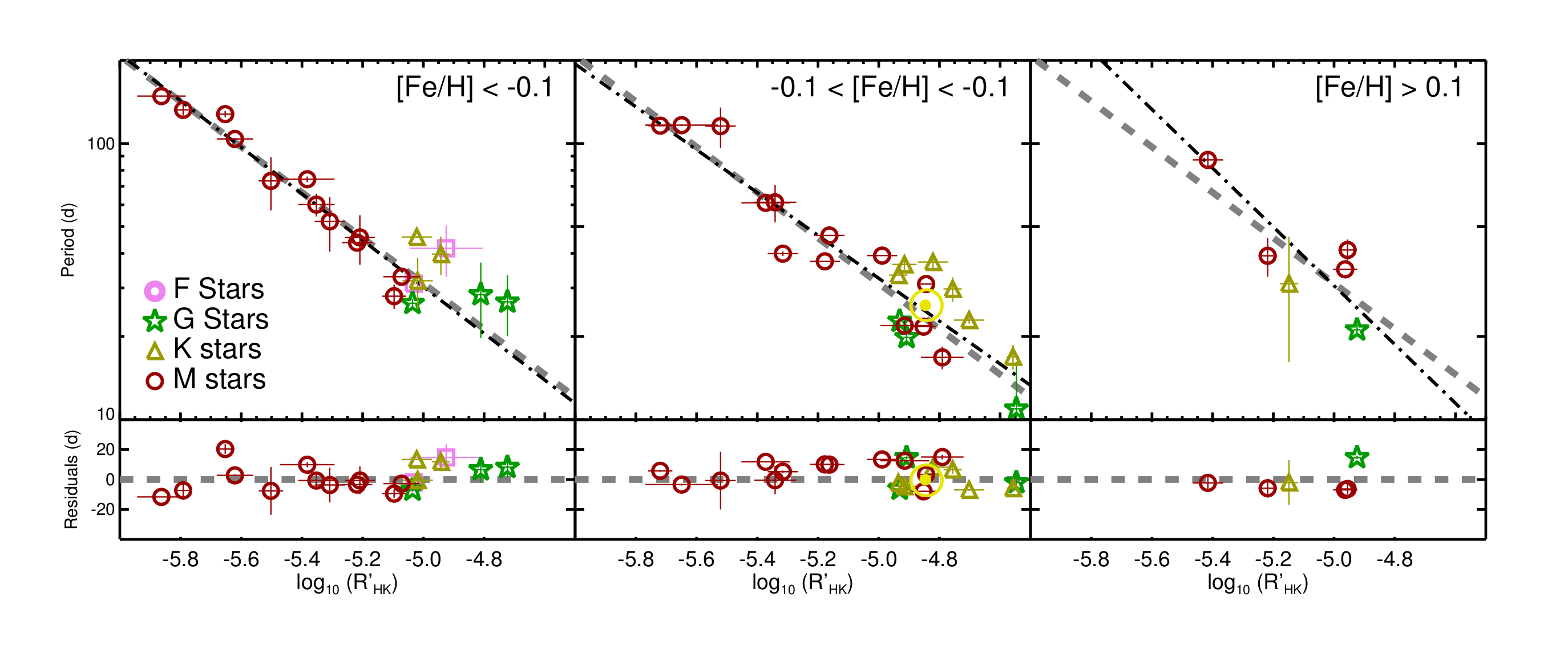}
\caption{Rotation periods versus chromospheric activity level  $\log_{10}(R'_\textrm{HK})$  in three metallicity intervals. The grey dashed line shows the fit to the full dataset, while the black dashed lines show  individual fits for each metallicity subset.} 
\label{period_met}
\end{figure*}

In principle metallicity may have an effect on the scatter seen in Figure~\ref{period_rhk}. In order to establish whether or not this is the case, we studied   the rotation period--chromospheric level relation for three  different metallicity ranges.  In Fig.~\ref{period_met} we display this relation for stars binned in each of these metallicity ranges. The plots show  the absence of  any clear dependence in the relatively small  metallicity range of our stars. The metallicities used can be seen in Table~\ref{tab:Data_Stars} and have been taken from \citet{Sousa2008},\citet{Santos2013},\citet{Tsantaki2013},\citet{RamirezAllende2013} and \citet{Neves2014}.

The Ca II H\&K and $H\alpha$ lines have proved to be  good indicators in the search for rotational modulation. Other lines could also correlate well with rotation. As shown by \citet{GomesdaSilva2011} and \citet{Robertson2015}, the Na I line, which should be included in future studies, also appears to be efficient at detecting stellar rotation periods.

\section{Implications for planet-hosting candidates}

Previous studies of rotational modulation  have questioned the existence of some planets through the  use of periodograms of time-series measurements of activity indicators \citep[e.g.][]{Robertson2014}. We have examined the orbital periods of   exoplanet claims for  our sample of stars  in the light of the rotational periods that we have measured and found   periodic signals in the activity time-series of several stars  that are very similar to the orbital periods of planet candidates previously reported in the literature. We comment below on these,  stars which are also listed in Table~\ref{table_danger}:

{\bf GJ 581}: this star hosts a  multi-planetary system, the real existence of some of the planets being questioned from the analysis of chromospheric activity time series.  It  was originally claimed to have 6 planets: planet b at $\sim5.4 d$ \citep{Bonfils2005}, planet c at $\sim12.9 d$  and d at $\sim66.6 d$ \citep{Udry2007}, planet e at $\sim3.1 d$ \citep{Mayor2009}, and planets f $\sim436 d$ and g at $\sim36 d$ \citep{Vogt2010}. Planets d and g have been challenged  as artefacts of stellar rotation \citep{Robertson2014}. Our own analysis of time series using HARPS spectra gives a rotation period of $\sim132~d$ for this star. This is about twice and 4 times the orbital periods of planet d and g, respectively (see Table~\ref{table_danger}). \citet{Boisse2011} explains the possible origin of this kind of signals as an effect of the latitude of the stellar structures and the line of sight. Our independent  measurement of rotation period adds  support to the conclusion of  \citet{Robertson2014}.

{\bf GJ 667C}: this is a 5-planet candidate system: planets b at $\sim7.2 d$ and c at $\sim28.1 d$ \citep{Bonfils2013}, and planets d at $\sim91.6 d$, e at $\sim62.2 d$ and f at $\sim39.0 d$ \citep{Gregory2012, AngladaEscude2013}. Recently, \citet{Robertson2014b} and \citet{Feroz2014} claimed that planets d and e are artefacts induced by stellar rotation. \citet{Delfosse2013} had already identified the $91 d$ signal as an alias of an  $\sim106 d$ activity signal.  Planet d, at $91\ d$, is  relatively close to our rotational period of $103 d$, but we also found a power excess at $\sim91 d$ in the $S_{\rm MW}$ periodogram, supporting the idea that the the signal is in fact an alias of the stellar rotation. For planet e, the claimed    $62~d$ orbital period  is very  close to  $P_{\rm rot} \cdot 2/3$~\citep{Santos2014}, casting doubts on its existence.

{\bf GJ 676A}: this star has 4 detected planets: planet b at $\sim1050d$  \citep{Forveille2011}, planet c at $\sim4400 d$ , planet d at $\sim3.6 d$ and planet e at $\sim35.4 d$  \citep{AngladaEscude2012}. The latter, planet e, has an orbital period compatible with our measured stellar rotation period ($41.2 d$). This makes us suspect that it could be an activity-induced signal and not a real planet. 

{\bf GJ 832}: there are 2 potential planets: planet b at $\sim3416d$ \citep{Bailey2009} and planet c at $\sim35.7d$ \citep{Wittenmyer2014}. The latter has an orbital period that falls inside the error bar of our value of stellar rotation period of $45.7d$, being really close to the Ca II H\&K period measurement. \citet{Bonfils2013} also found the signal of planet c, but noticed a power excess at $\sim 40$ days in the bisector periodogram. Further investigation is needed in order to confirm or reject the existence of planet c.

{\bf HD 40307}:After \citet{Mayor2009b} initial claim of 3 super-Earths at $40.3, 9.6$ and $20.4 d$, \citet{Tuomi2013} expanded the system to six planets, planet e being a small super-Earth at $\sim$34.6 day orbital period. This period coincides within the uncertainties with our stellar rotation detection ($31.8d$) and therefore the existence of the planet is in question.
 
{\bf HD 41248}: \citet{Jenkins2013} claimed the detection of two super-earths, at  periods $\sim$18 and $\sim$25 days. However, \citet{Santos2014} attributed these signals to the rotation period of the star at $\sim$25~d and to differential rotation ($\sim$18 days) across different latitudes. Our analysis of time series of activity indicators confirm the stellar rotation period of $\sim$25~d and therefore adds support to the doubts raised  concerning the existence of at least one of these planets.

\begin {table*}
\begin{center}
\caption {Planet candidates with orbital periods close to  rotation periods of  stars in the sample} \label{tab:title} 
\begin{tabular} { l l l l l l}    \hline
    Star  & Rot. Period & Planet & Orb. Period & Semi Amp.  & Ref \\ 
             & [d] &   & [d] & [m/s]  & \\ \hline
	
	GJ581        & 132.5 $\pm$ 6.3  & d & 66.64  $\pm$ 0.08   & 2.63 & 1,8  \\ 
	             &                  & g & 36.65  $\pm$ 0.52   & 1.5  & 2,8 \\ 
	GJ667C       & 103.9 $\pm$ 0.7  & d & 91.61  $\pm$ 0.89   & 1.52 & 3,9 \\ 
	             &                  & e & 62.2   $\pm$ 0.55   & 0.92 & 3,9 \\ 	
	GJ676A       & 41.2 $\pm$ 3.8   & e & 35.7   $\pm$ 0.07   & 2.62 & 4 \\ 	
	GJ832        & 45.7 $\pm$ 9.3   & c & 35.68  $\pm$ 0.03   & 1.6  & 5 \\ 
	HD40307      & 31.8  $\pm$ 6.7  & e & 36.64  $\pm$ 0.20   & 0.85 & 6  \\ 
	HD41248      &  26.4 $\pm$ 1.1  & b & 18.36  $\pm$ 0.07   & 2.93 & 7,10  \\
	             &                  & c & 25.65  $\pm$ 1.13   & 1.84 & 7,10  \\ \hline

\label{table_danger}
\end{tabular}      
\end{center}
\begin{flushleft}
\textbf{Ref:} 1 - \citet{Udry2007},
2 - \citet{Vogt2010},
3 - \citet{AngladaEscude2013}, 
4 - \citet{AngladaEscude2012}, 
5 - \citet{Wittenmyer2014}, 
6 - \citet{Tuomi2013}
7 - \citet{Jenkins2013}
8 - \citet{Robertson2014}
9 - \citet{Robertson2014b},
10 - \citet{Santos2014}
\end{flushleft}

\end {table*}

\section[]{Conclusions}

Using more than 6400 high resolution spectra from the HARPS public database we have analysed the time variability of the Ca II H\&K and H$\alpha$ lines in a sample of 48 late-type main sequence stars,  of which 29 are M-dwarfs.

We have been able to derive consistent periodicities in the time series of these activity indicators for relatively quiet stars ($\log_{10}(R'_\textrm{HK}) < -4.5$) of spectral types late F to mid M.
From these periodic signals we infer stellar rotation periods for the full sample of 48 stars. The derived periods cover the range 10--150~d. The average uncertainty of our period determinations is less than $\sim$10$\%$ of the measured period, being $\sim$5$\%$ the median value of the error distribution.
The rotation periods inferred from the Ca II H $\&$ K and the H$\alpha$ indexes are consistent within these uncertainties and compatible with periods inferred from photometric modulations reported in the literature.

We have investigated the correlation between the chromospheric emission, represented by the $\log_{10}(R'_\textrm{HK}$) index, and the derived spectroscopic periods, and we have obtained a linear relationship between this index and the logarithm of the rotation period, which for M dwarfs is valid in the range $\log_{10}(R'_\textrm{HK})\sim$ --4.5 to $\sim$ --5.8, and independent of metallicity in the range $ -0.3$ \textless [Fe/H] \textless 0.3. This relationship allows us to predict the rotational period of an M-dwarf star from spectroscopic time-series measurements. F-, G- and K-type stars in our sample also verify this relationship, although we have been able to explore a much limited range of chromospheric activity for these stars.

The radial velocity jitter caused by stellar chromospheric activity is a major concern in the search for terrestrial planets using RV surveys.  Among other effects, the signal induced by stellar rotation could easily be mistaken for the detection of low-mass planetary signals.  Some of our measured rotation periods are indeed similar to orbital periods of  candidate planets reported in the literature, casting some doubts on the existence of several of these planets. New generations of spectrographs will be able to measure radial velocities with a precision of a few $cm s^{-1}$, and a limiting factor for the detection of a terrestrial planet will probably be the activity of the stars. We have shown that the signal induced by stellar rotation can be systematically detected using spectral signatures present in the same high-quality spectra required for terrestrial planet  detection in RV surveys. Further studies are needed to characterize and  disentangle  stellar activity-induced signals from terrestrial planet signals  in time series of   high-precision RV measurements. 

\section*{Acknowledgements}

This work has been partially financed by the Spanish Ministry project AYA2011-26244.
J.I.G.H. also acknowledges the support from Spain's Ministry of Economy and 
Competitiveness (MINECO) under the 2011 Severo Ochoa Program 
MINECO SEV-2011-0187. This work is based on data obtained ARPS public database at the European Southern Observatory (ESO). This research has made extensive use of the SIMBAD database, operated at CDS, Strasbourg, France and NASA�s Astrophysics Data System. We are grateful to all the observers of the following ESO projects, whose data we are using 60.A-9036, 072.C-0096, 073.C-0784, 073.D-0038, 073.D-0578, 074.C-0012, 074.C-0364, 074.D-0131, 075.D-0194, 076.C-0878, 076.D-0130, 076.C-0155, 077.C-0364, 077.C-0530, 078.C-0044, 078.C-0833, 078.D-0071, 079.C-0681, 079.C-0927, 079.D-0075, 080.D-0086, 081.C-0148, 081.D-0065, 082.C-0212, 082.C-0308, 082.C-0315, 082.C-0718, 083.C-1001, 083.D-0040, 084.C-0229, 085.C-0063, 085.C-0019, 085.C-0318,
086.C-0230, 086.C-0284, 087.C-0368, 087.C-0831, 087.C-0990, 088.C-0011, 088.C-0323, 088.C-0353, 088.C-0662, 089.C-0050, 089.C-0006, 090.C-0421, 089.C-0497, 089.C-0732, 090.C-0849, 091.C-0034, 091.C-0866, 091.C-0936,  091.D-0469, 180.C-0886, 183.C-0437, 183.C-0972, 188.C-0265, 190.C-0027, 191.C-0505, 191.C-0873, 282.C-5036.

\bibliography{RHK_ref}

\begin {table*}
\begin{center}
\caption {Data for each star} \label{tab:Data_Stars} 
  \begin{tabularx}{\linewidth}{ l l l l  l  l  l l l l l l l l l l l l l}
    \hline
Name & SpTp$^{\ast}$ & N.Spec & TSp$^{\star}$ & $\textless S/N \textgreater$  & m$_\textrm{B}$ & m$_\textrm{V}$ & B-V & $S$ & $S_{phot}$ & $S_{R}+S_{V}$ & [Fe/H] & Dist. (pc)  & Ref. \\ \hline
HD25171 & F8 & 30 & 3597 & 44 & 8.29  &  7.78  &  0.51  &  0.13  &  0.07  &  7.47  &   -0.11 & 55.0 $\pm$ 1.4 & 1, 5, 10\\
HD1581 & F9 & 499 & 1167 & 125 & 4.80  &  4.23  &  0.57  &  0.15  &  0.09  &  231.62  &  -0.18 & 8.6 $\pm$ 0.1 & 1, 6, 11\\
HD1388 & G0 & 90 & 1250 & 235 & 7.09  &  6.50  &  0.59  &  0.15  &  0.07  &  28.95  &  -0.01 & 27.2 $\pm$ 0.4 & 1, 6, 12\\
HD41248 & G1 & 203 & 1843 & 78 &  9.42  &  8.81  &  0.61  &  0.16  &  0.08  &  2.57 &  -0.37 & 52.4 $\pm$ 1.9 & 1, 6, 10 \\
HD134060 & G1 & 110 & 1503 & 178 & 6.91  &  6.29  &  0.62  &  0.13  &  0.06  &  42.24  &  0.14 & 24.2 $\pm$ 0.3 & 1, 6, 10\\
HD30495 & G2 & 86 & 864 & 175 &  6.14  &  5.50  &  0.64  &  0.29 &  0.05  &  53.50  &  0.00 & 13.3 $\pm$ 0.1 & 1, 7, 10\\
HD1320 & G2 & 21 & 3230 & 135 & 8.63  &  7.98  &  0.65  &  0.22  &  0.09  &  5.77  &  -0.27 & 36.4 $\pm$ 0.9 & 1, 6, 12\\
HD2071 & G2 & 49 & 3700 & 170 &  7.95  &  7.27  &  0.68  &  0.18  &  0.08  &  10.86  &  -0.09 & 27.2 $\pm$ 0.5 & 1, 6, 10\\
HD63765 & G5 & 48 & 2302 & 137 & 8.85  &  8.10  &  0.75  &  0.23  &  0.05  &  5.65  &  -0.16 & 33.3 $\pm$ 0.7 & 1, 6, 13\\
Corot-7 & K0 & 164 & 1437 & 40 & 12.52  &  11.67  &  0.85  &  0.33  &  0.03  &  0.16  &  0.02 & 150.0 $\pm$ 20.0 & 2, 5, 14\\
HD224789 & K1 & 33 & 2833 & 113 & 9.12  &  8.24  &  0.88  &  0.54  &  0.05  &  2.46  &  -0.04  & 29.9 $\pm$ 0.6 & 1, 8, 10\\
HD77338 & K1 & 38 & 3205 &  80 & 9.47  &  8.59  &  0.88  &  0.17  &  0.04  &  1.99  &  0.28 & 39.2 $\pm$ 1.6 & 1, 5, 10\\
HD4628 & K2 & 164 & 1719 & 136 & 6.64  &  5.74  &  0.90  &  0.23  &  0.03  &  28.10  &  -0.31 & 7.5 $\pm$ 0.1 & 1, 7, 11, 15\\
HD176986 & K3 & 144 & 3585 & 99 & 9.39  &  8.45  &  0.94  &  0.27  &  0.05  &  2.39  &  0.03 & 26.4 $\pm$ 0.7 & 1, 8, 10\\
HD40307 & K3 & 442 & 3810 & 162 & 8.10  &  7.15  &  0.95  &  0.20  &  0.03  &  8.36 &  -0.36 & 13.0 $\pm$ 0.1 & 1, 8, 16\\
HD104067 & K3 & 86 & 2270 & 142 & 8.90  &  7.92  &  0.98  &  0.33  &  0.03  &  3.95  &  -0.04 & 21.1 $\pm$ 0.4 & 1, 8, 16\\
HD215152 & K3 & 265 & 3805 & 122 & 9.12  &  8.13  &  0.99  &  0.26  &  0.04  &  3.12 &  -0.08 & 21.5 $\pm$ 0.5 & 1, 8, 10\\
HD125595 & K4 & 137 & 2075 & 57 & 10.13  &  9.03  &  1.10  &  0.49  &  0.04  &  1.02 &  0.10 & 28.0 $\pm$ 0.8 & 1, 5, 12\\
HD85512 & K6 & 777 & 2336 & 128 & 8.83  &  7.65  &  1.18  &  0.45  &  0.04  &  2.63  &  -0.26 & 11.2 $\pm$ 0.1 & 1, 8, 16\\
GJ676A & M0 & 88 & 2263 & 44 & 11.03  &  9.59  &  1.44  &  1.40  &  0.06  &  0.29  &  0.26 & 16.5 $\pm$ 0.5 & 1, 9, 16\\
GJ1 & M0 & 42 & 2063 & 108 & 10.02  &  8.56  &  1.46  &  0.60  &  0.05  &  1.19  &  -0.45 & 4.3 $\pm$ 0.1 & 1, 9, 16\\
GJ536 & M1 & 104 & 3628 & 48 & 11.18  &  9.71  &  1.47  &  0.85  &  0.04  &  0.25  &  -0.14 & 10.0 $\pm$ 0.2 & 1, 9, 16\\
GJ436 & M1 & 152 & 1526 &  33 & 12.06  &  10.59  &  1.47  &  0.65  &  0.05  &  0.12 &  -0.03 & 10.1 $\pm$ 0.3 & 1, 9 ,10\\
GJ846 & M1 & 55 & 2345 & 70 & 10.62  &  9.15  &  1.47  &  1.76  &  0.05  &  0.47  &  0.01 & 10.2 $\pm$ 0.2 & 1, 9, 16\\
GJ205 & M1 & 76 & 1400 & 157 & 9.44  &  7.97  &  1.47  &  1.45  &  0.04  &  2.13  &  0.19 & 5.7 $\pm$ 0.1 & 1, 9, 16\\
GJ382 & M1 & 32 & 1187 & 68 & 10.76  &  9.26  &  1.49  &  1.91  &  0.04  &  0.68  &  0.02 & 7.9 $\pm$ 0.2 & 1, 9, 16, 17\\
GJ163 & M1 & 165 & 3704 & 26 & 13.30  &  11.81  &  1.49  &  0.55  &  0.04  &  0.07  &  0.07 & 15.0 $\pm$ 0.5 & 1, 9, 16\\
GJ526 & M1 & 29 & 1790 & 96 & 9.92  &  8.43  &  1.49  &  0.75  &  0.04  &  1.31 &  -0.22 & 5.4 $\pm$ 0.1 & 1, 9, 16\\
GJ514 & M1 & 114 & 3634 & 73 & 10.52  &  9.03  &  1.49  &  1.12  &  0.06  &  0.39  &  -0.16 & 7.7 $\pm$ 0.1 & 1, 9, 16\\
GJ877 & M2 & 46 & 2973 & 43 & 11.87  &  10.38  &  1.50  &  0.33  &  0.05  &  0.16 &  0.00 & 8.6 $\pm$ 0.1 & 1, 9, 16\\
GJ849 & M2 & 42 & 1893 & 46 & 11.87  &  10.37  &  1.50  &  0.94  &  0.04  &  0.23 & 0.24 & 8.6 $\pm$ 0.2 & 4, 9 ,16\\
GJ588 & M2 & 188 & 3271 & 55 & 10.81  &  9.31  &  1.50  &  0.85  &  0.04  &  0.32  &  0.06 & 5.9 $\pm$ 0.1 & 1, 9, 16\\
GJ832 & M2 & 53 & 3134 & 97 & 10.18  &  8.67  &  1.50  &  0.75  &  0.04  &  0.88  &  -0.17 & 5.0 $\pm$ 0.1 & 1, 9, 16\\
GJ3470 & M2 & 78 & 1434 & 15 & 13.77$^{\ddagger}$ &  12.27  &  1.50  &  1.62  &  0.03  &  0.02 &  0.08 & 28.8 $\pm$ 2.6 & 3, 5, 19\\
GJ880 & M2 & 77 & 3636 & 79 & 10.14  &  8.64  &  1.51  &  1.56  &  0.07  &  0.49  &  0.03 & 6.9 $\pm$ 0.1 & 1, 9, 16\\
GJ433 & M2 & 79 & 2490 & 58 & 11.32  &  9.81  &  1.51  &  0.57  &  0.06  &  0.29  &  -0.17 & 8.9 $\pm$ 0.2 & 1, 9, 16\\
GJ752A & M2 & 73 & 3609 & 77 &  10.63  &  9.12  &  1.52  &  1.09  &  0.02  &  0.48  &  0.05 & 5.9 $\pm$ 0.1 & 1, 9, 16\\
GJ701 & M2 & 68 & 3302 & 64 & 10.88  &  9.36  &  1.52  &  0.50  &  0.04  &  0.31  &  -0.27 & 7.8 $\pm$ 0.1 & 1, 9, 16\\
GJ358 & M2 & 27 & 1923 & 47 & 12.21  &  10.69  &  1.52  &  3.03  &  0.04  &  0.19  &  -0.01 & 9.5 $\pm$ 0.2 & 1, 9, 16\\
GJ176 & M3 & 68 & 2146 & 55 & 11.49  &  9.95  &  1.54  &  1.56  &  0.04  &  0.27 &  -0.01 & 9.3 $\pm$ 0.3 & 1, 9, 16\\
GJ876 & M3 & 225 & 3089 & 51 & 11.75  &  10.19  &  1.56  &  0.77  &  0.06  &  0.25  &  0.14 & 4.7 $\pm$ 0.1 & 1, 9, 18\\
GJ674 & M3 & 137 & 3580 & 81 & 10.97  &  9.41  &  1.57  &  1.56  &  0.04  &  0.56  &  -0.23 & 4.6 $\pm$ 0.1 & 1, 9, 16\\
GJ667C & M3 & 167 & 3073 & 48 & 11.79  &  10.22  &  1.57  &  0.60  &  0.03  &  0.20  &  -0.50 & 7.1 $\pm$ 0.2 & 4, 9, 12\\
GJ273 & M3 & 183 & 2349 & 60 & 11.44  &  9.87  &  1.57  &  0.70  &  0.04  &  0.30 & -0.01 & 3.8 $\pm$ 0.1 & 1, 9, 16\\
GJ357 & M3 & 47 & 2521 & 34 & 12.48  &  10.91  &  1.57  &  0.83  &  0.04  &  0.11  &  -0.30 & 9.0 $\pm$ 0.2 & 1, 9, 16\\
GJ581 & M4 & 251 & 2909 & 43 & 12.17  &  10.57  &  1.60  &  0.45  &  0.04  &  0.15  &  -0.20 & 6.2 $\pm$ 0.1 & 4, 9, 16\\
GJ699 & M5 & 211 & 2234 & 54 & 11.24  &  9.51  &  1.73  &  0.68  &  0.03  &  0.23  & -0.51 & 1.8 $\pm$ 0.1 & 1, 9 ,16\\
GJ551 & M6 & 222 & 3515 & 23 & 12.97  &  11.10  &  1.87  &  8.06  &  0.03  &  0.03  & 0.00 & 1.3 $\pm$ 0.1 & 4, 9, 20\\ \hline

\label{table_data}
\end{tabularx}  
\end{center}
\begin{flushleft}
\textbf{References for distance:} 1 - \citet{vanLeeuwen2007}, 
2 - \citet{Leger2009}, 
3 - \citet{Biddle2014}, 
4 - \citet{Laurie2014}. \\
\textbf{References for metallicity:} 
5 - \citet{Santos2013}, 
6 - \citet{Sousa2008}, 
7 - \citet{RamirezAllende2013}, 
8 - \citet{Tsantaki2013}, 
9 - \citet{Neves2014}.\\
\textbf{References for magnitudes:}
10 - \citet{Hog2000}$^{\dagger}$,
11 - \citet{Ducati2002},
12 - \citet{Mermilliod1986},
13 - \citet{Cousins1962},
14 - \citet{Leger2009}
15 - \citet{vanBelle2009},
16 - \citet{Koen2010},
17 - \citet{Kiraga2012},
18 - \citet{Landolt2009},
19 - \citet{Rapaport2001},
20 - \citet{JaoHenry2014}. \\
$^{\ast}$ Approximate spectral type based in the colour index.\\
$^{\star}$ Time span of the observations in days.\\
$^{\dagger}$ Tycho magnitudes converted to Johnson magnitudes. \\
$^{\ddagger}$ No available Johnson $m_{B}$ in the literature. Value calculated using the $(B-V)$ value of the star with the closest spectrum using \citet{Reid1997} as guide. \\

\end{flushleft}
\end {table*}
\label{lastpage}

\end{document}